\documentclass[arxiv]{agn_article}
\usepackage{multirow}
\usepackage{agn_all}
\usepackage{float}
\graphicspath{{images/}}

\newcommand{\comsol}{\textit{Comsol Multiphysics}\textsuperscript{\textregistered}}
\newcommand{\mathematica}{\textit{Mathematica}}
\providecommand{\kappabar}{\overline{\kappa}}
\providecommand{\gammabar}{\overline{\gamma}}

\defineauthor{madeo}{Angela~Madeo}{bmsd}{angela.madeo@tu-dortmund.de}
\defineauthor{rizzi}{Gianluca~Rizzi}{bmsd}{gianluca.rizzi@tu-dortmund.de}

\begin{document}

\title{Modeling a labyrinthine acoustic metamaterial through an inertia-augmented relaxed micromorphic approach}
\knownauthors[vossd]{vossd,rizzi,neff,madeo}
\date{\today}
\maketitle

\begin{abstract}
\noindent
We present an inertia-augmented relaxed micromorphic model that enriches the relaxed micromorphic model previously introduced by the authors via a term $\Curl \dot{P}$ in the kinetic energy density.
This enriched model allows us to obtain a good overall fitting of the dispersion curves while introducing the new possibility of describing modes with negative group velocity that are known to trigger negative refraction effects.
The inertia-augmented model also allows for more freedom on the values of the asymptotes corresponding to the cut-offs.
In the previous version of the relaxed micromorphic model, the asymptote of one curve (pressure or shear) is always bounded by the cut-off of the following curve of the same type.
This constraint does not hold anymore in the enhanced version of the model.
While the obtained curves' fitting is of good quality overall, a perfect quantitative agreement must still be reached for very small wavelengths that are close to the size of the unit cell.
\end{abstract}

\textbf{Key words:} metamaterials, metastructure, inertia-augmented, relaxed micromorphic model, anisotropy, dispersion curves, band-gap, parameters identification, generalized continua
\\[.65em]
\noindent\textbf{AMS 2010 subject classification:
    74A10, 
 	74B05, 
 	74J05, 
 	74M25  
 }

 {\parskip=-0.4mm\tableofcontents}

\section{Introduction}\label{sec:introduction}
Metamaterials are materials whose mechanical properties go beyond those of classical materials thanks to their heterogeneous microstructure.
They can show unusual static/dynamic responses such as negative Poisson’s ratio \cite{lakes1987foam}, twist or bend in response to being pushed or pulled \cite{frenzel2017three,rizzi2019identification}, band-gaps \cite{liu2000locally,wang2014harnessing,bilal2018architected,celli2019bandgap}, cloaking \cite{buckmann2015mechanical,misseroni2016cymatics}, focusing \cite{guenneau2007acoustic,cummer2016controlling}, channeling \cite{kaina2017slow,tallarico2017edge}, negative refraction \cite{zhu_negative_2014,kaina_negative_2015,willis_negative_2016}, etc.
The working frequency of each metamaterial strongly depends on the characteristic size and the geometry of the underlying unit cell, as well as on the choice of the base material.
In this paper, we present a labyrinthine metamaterial that, thanks to the use of a polymeric based material and an optimized distribution of mass inside the unit cell (see Figure \ref{fig:ChartesGeometry}), gives rise to a wide acoustic band-gap with characteristic unit cell’s size of the order of centimeters.

The direct finite element modeling of structures build up with this labyrinthine metamaterial is unfeasible due to the extremely tight meshing that would be needed to correctly cover the narrow strips of material inside each unit cell.
It is thus apparent the need for a homogenized model to use this type of very promising metamaterials in actual engineering designs.
Various homogenization techniques have been developed with the purpose of providing rigorous predictions of the macroscopic metamaterial’s mechanical response when the properties of the base materials and their spatial distribution are known.
These homogenization approaches have been shown to be useful in describing the overall behavior of metamaterials in the static and quasi-static regimes \cite{bensoussan2011asymptotic,sanchez1980non,allaire1992homogenization,milton2002theory,hashin1963variational,willis1977bounds,pideri1997second,bouchitte2002homogenization,camar2003determination,suquet1985elements,miehe1999computational,geers2010multi,hill1963elastic} as well as, more recently, in the dynamic regime \cite{bacigalupo2014second,chen2001dispersive,boutin2014large,craster2010high,andrianov2008higher,hu2017nonlocal,willis2009exact,willis2011effective,willis2012construction,srivastava2014limit,sridhar2018general,srivastava2017evanescent}.
However, these models are often unsuited to deal with finite-size metamaterials, because they are based on upscaling techniques valid for unbounded media.
Because of that, finite-size metamaterials’ structures are mostly investigated via Finite Element simulations which are performed using directly the microstructured material, e.g.\ \cite{krushynska2017coupling}.
The downside of this approach is that the computational cost quickly becomes unsustainable (especially for unit cells as the one presented in this paper), although the propagation patterns obtained are very accurate.
This heavily limits the possibility of exploring large-scale or very convoluted geometric meta-structures.

To overcome this problem and open up the possibility of designing complex meta-structures using the metamaterial presented in this paper as a basic building block, we propose to use an inertia-augmented relaxed micromorphic model.
This model is based on the relaxed micromorphic model that we previously established \cite{neff_unifying_2014,neff_relaxed_2015,dagostino_effective_2020,aivaliotis_microstructure-related_2019,aivaliotis_frequency_2020} and has been augmented with a new inertia term accounting for coupled space-time derivatives of the micro-distortion tensor.
The relaxed micromorphic model has extensively proven its efficacy in describing the broadband behavior of many infinite and finite-size metamaterials \cite{aivaliotis_microstructure-related_2019,aivaliotis_frequency_2020,rizzi_exploring_2021,rizzi2021boundary,rizzi2021metamaterial} and is extended in this paper so as to be able to account for negative group velocity which was not the case before.
We will show that the proposed model is able to describe well the labyrinthine metamaterial’s response for a large range of frequencies (going beyond the first band-gap) and wave numbers (approaching the size of the unit cell) and for all directions of propagation with a limited number of frequency- and scale-independent constitutive parameters.
The new inertia-augmented term will be shown to trigger modes with negative group velocities that are known to be associated with negative refraction phenomena.
The results presented in this paper will allow us to shortly present new designs of finite-size labyrinthine metamaterials’ structures that can control elastic energy in the acoustic regime for eventual subsequent re-use.

\subsection{A Polyethylene-based metamaterial for acoustic control}\label{sec:unitCell}
In this section, we present a new unit cell's design that gives rise to a metamaterial for acoustic control.
This unit cell is designed to achieve a band-gap at relatively low frequencies ($600-2000~\si{\Hz}$) so that application for acoustic control can be targeted.
The unit cell considered is made out of polyethylene, cf.\ Table~\ref{tab:numericalValuesMacro}.
Compared to Aluminium or Titanium, that we used for the metamaterials studied in \cite{rizzi_exploring_2021,rizzi2021boundary,rizzi2021metamaterial}, Polyethylene gives rise to lower wave speeds, thus allowing band-gap phenomena to appear at lower frequencies.

\begin{figure}[ht!]
    \begin{minipage}{0.49\textwidth}
	    \centering
	    \includegraphics[width=\textwidth, trim = 7cm 11cm 7cm 11cm, clip]{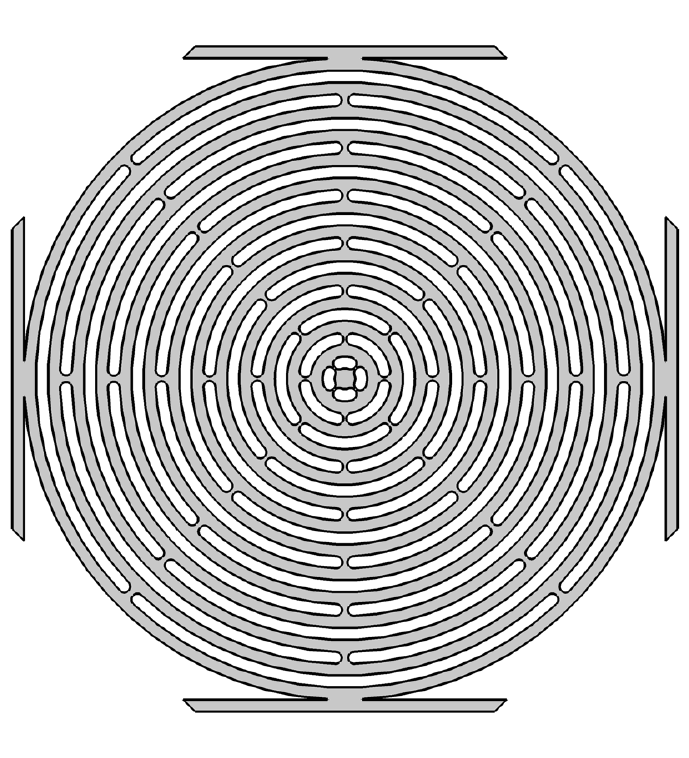}
	\end{minipage}
	\hfill
	\begin{minipage}{0.49\textwidth}
        \centering
        \begin{tikzpicture}
            \node (1) [anchor=south west, inner sep=0pt] {\includegraphics[width=\textwidth, trim = 6.7cm 11cm 6.7cm 11cm, clip]{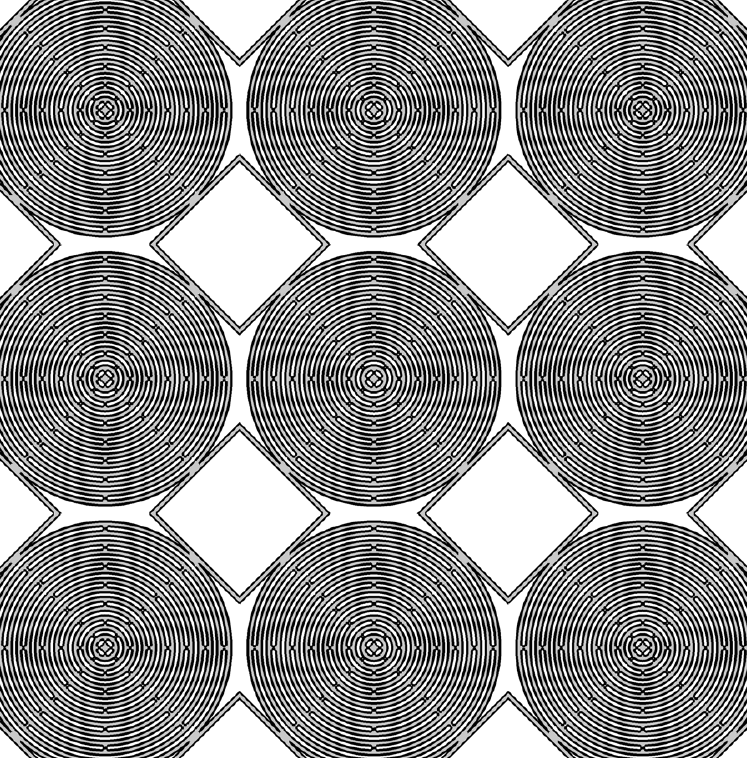}};
            \draw[draw=red,dashed,thick] (2.55,2.6) rectangle ++ (2.85,2.85);
            \node[red] at (5.15,5.65) {$a$};
	    \end{tikzpicture}
    \end{minipage}
	\caption{Geometry of the unit cell. Left: details of one unit cell (rotated by 45 degrees) showing the tetragonal symmetry. For $a=20\.\si{\milli\m}$ as the size of the unit size we consider later, both the bars and holes have a thickness of $0.4\.\si{\milli\m}$ each. Right: A 3x3 section of the metamaterial made up of this unit cell (red dashed square).}
	\label{fig:ChartesGeometry}
\end{figure}

\renewcommand{\arraystretch}{1.5}
\begin{table}[ht!]
    \centering
    \begin{tabular}{c|c|c|c|c|c}
        $a$  & $\rho$  & $\overline{\kappa}_{\rm B}$ & $\mu_{\rm B}$ & $c_{\rm p,B}$ & $c_{\rm s,B}$
        \\ \hline
        $[\si[per-mode = symbol]{\mm}]$ & $[\si[per-mode = symbol]{\kg\per\cubic\m}]$ & $[\si{\MPa}]$ & $[\si{\MPa}]$ & $[\si[per-mode = symbol]{\m\per\s}]$ & $[\si[per-mode = symbol]{\m\per\s}]$
        \\ \hline
        20 & 900 & 3160 & 262 & 1950 & 540
    \end{tabular}
    \caption{
    Size, bulk material constants and wave speed for polyethylene unit cell where $\overline{\kappa}_{\rm B}$ is the plane strain bulk modulus.\protect\footnotemark
    }
    \label{tab:numericalValuesMacro}
\end{table}
\footnotetext{The relations between the wave speed and the elastic constants is $c_{\rm p}=\sqrt{\frac{\overline{\kappa}_{\rm B}+\mu_{\rm B}}{\rho}}=\sqrt{\frac{\lambda_{\rm B}+2\mu_{\rm B}}{\rho}}$ and $c_{\rm s}=\sqrt{\frac{\mu_{\rm B}}{\rho}}$, while the plane strain bulk modulus is $\overline{\kappa}_{\rm B}=\lambda_{\rm B}+\mu_{\rm B}$.}
A further lowering of the band-gap is obtained through the adoption of a labyrinth-type geometry, cf.\ Figure~\ref{fig:ChartesGeometry}. This structure presents a tetragonal symmetry and thus features a reduced number of parameters with respect to a fully anisotropic system. The circular center of the unit cell is connected by thin bars allowing the heavier center to move easily, thus giving rise to local resonance phenomena of relatively low frequencies while additionally providing a very soft macro-material behaviour.

%
%
%
%
\section{Relaxed micromorphic modelling of finite-size metamaterials}\label{sec:const_law}

We briefly recall the weak and strong form of the governing equations of both classical Cauchy and relaxed micromorphic continua.
The Lagrangian $\mathcal{L}_{\rm c}$ for the classical Cauchy model is 
\begin{align}
    \mathcal{L}_{\rm c} \left(\nabla u,\dot{u}\right) =
    \frac{1}{2}\rho\.\norm{\dot{u}}^2
    -\frac{1}{2}\.\iprod{\mathbb{C}\.\sym\nabla u, \sym\nabla u}\,,
    \label{eq:class_energy}
\end{align}
where $u$ is the displacement field, $\iprod{ \cdot , \cdot }\col\Rnn\times\Rnn\to\R$ is the scalar product, $\rho$ is the apparent mass density, and $\mathbb{C}$ is the classical 4th order elasticity tensor.
The Lagrangian $\mathcal{L}_{\rm m}$ for the relaxed micromorphic model enhanced with the micro-inertia term Curl$\dot{P}$ is \cite{aivaliotis_microstructure-related_2019,aivaliotis_frequency_2020,rizzi_exploring_2021,rizzi2021boundary}\footnote{Here we target a complete expression of the Lagrangian that was not considered in \cite{aivaliotis_microstructure-related_2019,aivaliotis_frequency_2020,rizzi_exploring_2021,rizzi2021boundary} to show the effect on the predictability of the dispersion curves provided by each term.}
\begin{align}
\mathcal{L}_{\rm m} \left(\dot{u},\nabla \dot{u}, \dot{P}, \Curl \dot{P}, \nabla u, P, \Curl P \right) 
&=\dfrac{1}{2}\rho\.\norm{\dot{u}}^2
+\frac12\iprod{\mathbb{J}_{\rm m}\.\sym\dot P,\sym\dot P}+\iprod{\mathbb{J}_{\rm c}\.\skew\dot P,\skew\dot P}\label{eq:relax_energy}
\\
&\phantom{=}\ +\frac12\iprod{\mathbb{T}_{\rm e}\.\sym\nabla \dot{u},\sym\nabla \dot{u}}+\frac12\iprod{\mathbb{T}_{\rm c}\.\skew\nabla \dot{u},\skew\nabla \dot{u}}
\notag\\
&\phantom{=}\ +\frac12\iprod{\mathbb{M}_{\rm s}\.\sym\Curl\dot P,\sym\Curl\dot P}+\frac12\iprod{\mathbb{M}_{\rm a}\.\skew\Curl\dot P,\skew\Curl\dot P}
\notag\\
&\phantom{=}\ -\frac12\iprod{\mathbb{C}_{\rm e}\.\sym(\nabla u-P),\sym(\nabla u-P)}
\notag\\ 
&\phantom{=}\ -\frac12\iprod{\mathbb{C}_{\rm c}\.\skew(\nabla u-P),\skew(\nabla u - P)} -\frac12\iprod{\mathbb{C}_{\rm micro}\.\sym P,\sym P}
\notag\\
&\phantom{=}\ -\frac12\iprod{\mathbb{L}_{\rm s}\.\sym\Curl P,\sym\Curl P} -\frac12\iprod{\mathbb{L}_{\rm a}\.\skew\Curl P,\skew\Curl P},
\notag
\end{align}
where $u \in \mathbb{R}^{3}$ is the macroscopic displacement field, $P \in \mathbb{R}^{3\times 3}$ is the non-symmetric micro-distortion tensor, $\rho$ is the macroscopic apparent density, $\mathbb{J}_{\rm m}$, $\mathbb{J}_{\rm c}$, $\mathbb{T}_{\rm e}$, $\mathbb{T}_{\rm c}$, $\mathbb{M}_{\rm s}$, $\mathbb{M}_{\rm a}$, are 4th order micro-inertia tensors, and $\mathbb{C}_{\rm e}$, $\mathbb{C}_{\rm c}$, $\mathbb{C}_{\rm micro}$, $\mathbb{L}_{\rm s}$, and $\mathbb{L}_{\rm a}$ are 4th order elastic tensors (for more details see Appendix~\ref{app:SymSkewCoup}).\footnote{
The tensors $\mathbb{J}_{\rm m}$, $\mathbb{T}_{\rm e}$, $\mathbb{C}_{\rm e}$, $\mathbb{C}_{\rm micro}$, $\mathbb{L}_{\rm s}$ and $\mathbb{M}_{\rm s}$ have a minor symmetry $\left(\mathbb{A}_{\text{(ij)(kl)}}=\mathbb{A}_{\text{(ji)(kl)}}=\mathbb{A}_{\text{(ij)(lk)}}=\mathbb{A}_{\text{(ji)(lk)}}\right)$ while $\mathbb{J}_{\rm c}$, $\mathbb{T}_{\rm c}$, $\mathbb{C}_{\rm c}$, $\mathbb{L}_{\rm a}$ and $\mathbb{M}_{\rm a}$ has a minor anti-symmetry $\left(\mathbb{B}_{\text{[ij][kl]}}=-\mathbb{B}_{\text{[ji][kl]}}\right.$, $\mathbb{B}_{\text{[ij][kl]}}=-\mathbb{B}_{\text{[ij][lk]}}$, $\left.\mathbb{B}_{\text{[ij][kl]}}=\mathbb{B}_{\text{[ji][lk]}}\right)$, for instance we can write $\iprod{\mathbb{J}_{\rm m}\.\sym\dot P,\sym\dot P}+\iprod{\mathbb{J}_{\rm c}\.\skew\dot P,\skew\dot P}=\iprod{(\mathbb{J}_{\rm m}+\mathbb{J}_{\rm c})\.\dot P,\dot P}$.}
The action functional $\mathcal{A}_i$ of the considered continuum can be defined based on the Lagrangian function $\mathcal{L}_{i}$ and the external work $\mathcal{U}_{i}$ as 
\begin{align}
    \mathcal{A}_{\rm c}=
    \int\limits_{\mathclap{\Omega\times\left[0,T\right]}}\mathcal{L}_{\rm c}\,\dx\. \dt
    -\int_{0}^{T}\mathcal{U}_{\rm c}^{\rm ext}\,\dt
    \,,
    \quad\quad\quad\text{and}\quad\quad\quad
    \mathcal{A}_{\rm m}=
    \int\limits_{\mathclap{\Omega\times\left[0,T\right]}} \mathcal{L}_{\rm m}\,\dx\.\dt
    -\int_{0}^{T}\mathcal{U}_{\rm m}^{\rm ext}\, \dt
    \,,
    \label{eq:action_func}
\end{align}
for classical Cauchy and relaxed micromorphic media, respectively.
In equation \eqref{eq:action_func}, $\Omega\subset\R^3$ is the domain of the considered continuum in its reference configuration and $[0,T]$ is a time interval during which the deformation of the continuum is observed.
In the case of a conservative system, the virtual work of internal actions $\mathcal{W}_{i}^{\rm int}=\delta \mathcal{L}_{i}$ can be defined as the first variation of the Lagrangian function, while the virtual work of external actions $\mathcal{W}_{i}^{\rm ext}=\delta \mathcal{U}_{i}^{\rm ext}$ is given as the first variation of the external work. In formulas, we have
\begin{align}
    \delta\mathcal{A}_{\rm c}=\int_{0}^{T}\mathcal{W}_{\rm c}^{\rm int}\,\dt
    -\int_{0}^{T}\mathcal{W}_{\rm c}^{\rm ext}\, \dt
    \,,
    \quad\quad\quad\text{and}\quad\quad\quad
    \delta\mathcal{A}_{\rm m}=\int_{0}^{T}\mathcal{W}_{\rm m}^{\rm int}\,\dt
    -\int_{0}^{T}\mathcal{W}_{\rm m}^{\rm ext}\, \dt
    \,,
    \label{eq:first_var_A}
\end{align}
for classical Cauchy and relaxed micromorphic media, respectively.
In equation \eqref{eq:first_var_A}, the variation operator $\delta$ indicates that the variation must be taken with respect to the unknown kinematics fields ($u$ for Cauchy media and ($u,P$) for the relaxed micromorphic model).
Following classical variational calculus, the strong form of the bulk equations of motion and the Neumann boundary conditions for the Cauchy and the relaxed micromorphic model can be obtained via a least action principle stating that the first variation of the action functional must be vanishing.
In absence of external body loads, the application of a least-action principle to Cauchy and relaxed micromorphic models gives the following equilibrium equation \cite{rizzi_exploring_2021,aivaliotis_frequency_2020,neff_unifying_2014}
\begin{align}
    \rho\.\ddot{u} = \Div\sigma \,,
    \qquad\qquad
    \sigma
    \coloneqq
    \mathbb{C}\.\sym\nabla u\,,
\label{eq:equiCau}
\end{align}
for the classical Cauchy model, and
\begin{equation}
\rho\.\ddot{u} - \Div\sigmahat = \Div\widetilde{\sigma} \,,
\qquad
\sigmabar = \widetilde{\sigma} - s -\Curl m -\Curl\mhat \,,
\label{eq:equiMic}
\end{equation}
for the relaxed micromorphic model, where we set
\begin{align}
    \widetilde{\sigma}&\coloneqq
    \mathbb{C}_{\rm e}\.\sym(\nabla u-P) + \mathbb{C}_{\rm c}\.\skew(\nabla u-P)
    &\sigmahat&\coloneqq
    \mathbb{T}_{\rm e}\.\sym\nabla\ddot{u} + \mathbb{T}_{\rm c}\.\sym\nabla\ddot{u}\,,\notag\\
    \sigmabar&\coloneqq
    \mathbb{J}_{\rm m}\.\sym\ddot{P} + \mathbb{J}_{\rm c}\.\skew\ddot{P}\,,
    &s&\coloneqq
    \mathbb{C}_{\rm micro}\. \sym P \,,\label{eq:equiSigAll}\\
    m&\coloneqq
    \mathbb{L}_{\rm s}\.\sym\Curl P + \mathbb{L}_{\rm a}\.\skew\Curl P\,,
    &\mhat&\coloneqq
    \mathbb{M}_{\rm s}\.\sym\Curl\ddot{P} + \mathbb{M}_{\rm a}\.\skew\Curl\ddot{P} \,.
    \notag
\end{align}
The Neumann boundary condition for the classical Cauchy model are
\begin{align}
t_{\rm c} \coloneqq\. \sigma n=t_{\rm c}^{\rm ext},
\label{eq:traction_Cauchy}
\end{align}
with $t_{\rm c}^{\rm ext}$ as the externally applied traction vector and $n$ as the outward-pointing normal to the boundary, and for the relaxed micromorphic model
\begin{align}
t_{\rm m} \coloneqq \left(\widetilde{\sigma} + \sigmahat \right) n=t_{\rm m}^{\rm ext}
\qquad\text{and}\qquad
\tau \coloneqq \left(m + \mhat \right) \times n=\tau^{\rm ext} \,,
\label{eq:traction}
\end{align}
where $t_{\rm m}^{\rm ext}$ is the generalized traction vector, $\tau^{\rm ext}$ is the double traction second order tensor, and the cross product $\times$ is understood row-wise.
In absence of curvature terms ($\mathbb{L}_{\rm s}=\mathbb{L}_{\rm a}=\mathbb{M}_{\rm s}=\mathbb{M}_{\rm a}=0$), the boundary condition \eqref{eq:traction}$_2$ involving $\tau$ must not be assigned on the boundary.

As it is well-known, Dirichlet type boundary conditions can also be alternatively considered in the form $u=u^{\rm ext}$ for Cauchy media while $u=u^{\rm ext}$ and $P\times n=\phi^{\rm ext}$ for the relaxed micromorphic model where $n$ is the outward-pointing normal vector to the surface and $\phi^{\rm ext}$ is an assigned second order tensor.

\subsection{Tetragonal Symmetry / Shape of elastic tensors (in Voigt notation)}\label{sec:elasticTensors}
In the following equation \eqref{eq:micro_ine_1}, we report the elastic tensors expressed in Voigt notation for the tetragonal class of symmetry\footnote{The dimensions of the matrix representation of elastic and micro-inertia tensors are ${(\mathbb{C}_{\rm e},\mathbb{C}_{\rm micro},\mathbb{J}_{\rm m},\mathbb{T}_{\rm e},\mathbb{L}_{\rm s},\mathbb{M}_{\rm s})\in \mathbb{R}^{6\times 6}}$ and ${(\mathbb{C}_{\rm c},\mathbb{J}_{\rm c},\mathbb{T}_{\rm c},\mathbb{L}_{\rm a},\mathbb{M}_{\rm a})\in \mathbb{R}^{3\times 3}}$. In the previous works \cite{neff_identification_2020,dagostino_effective_2020,rizzi2021boundary}, the parameters in $\mathbb{J}_{\rm m}$ and $\mathbb{J}_{\rm c}$ were referred as $\eta_i$ or $\rho\.L_i^2$, and the parameters in $\mathbb{T}_{\rm e}$, $\mathbb{T}_{\rm c}$ as $\overline{\eta}_i$ or $\rho\.\overline{L}_i^2$.}, where only the parameters involved under the plane-strain hypothesis are explicitly presented. Thereby, the symbol $\star$ indicates that the specific entry do not intervene under the plane-strain hypothesis.\footnote{We retain the plane strain hypothesis in the remainder of the paper.}
The class of symmetry has been chosen accordingly with the symmetry of the unit cell presented in Figure~\ref{fig:ChartesGeometry}, under the assumption that the same class of symmetry applies both at the micro- and the macro-scale. 
\begin{align}
    \mathbb{C}_{\rm e}=
    &\begin{pmatrix}
    \kappa_{\rm e} + \mu_{\rm e}	& \kappa_{\rm e} - \mu_{\rm e}				& \star & \dots	& 0\\ 
    \kappa_{\rm e} - \mu_{\rm e}	& \kappa_{\rm e} + \mu_{\rm e} & \star & \dots & 0\\
    \star & \star & \star & \dots & 0\\
    \vdots & \vdots	& \vdots & \ddots &\\ 
    0 & 0 & 0 & & \mu_{\rm e}^{*}
    \end{pmatrix},
    &\mathbb{C}_{\rm micro}= 
    &\begin{pmatrix}
    \kappa_{\rm m} + \mu_{\rm m}	& \kappa_{\rm m} - \mu_{\rm m}				& \star & \dots	& 0\\ 
    \kappa_{\rm m} - \mu_{\rm m}	& \kappa_{\rm m} + \mu_{\rm m} & \star & \dots & 0\\
    \star & \star & \star & \dots & 0\\
    \vdots & \vdots	& \vdots & \ddots &\\ 
    0 & 0 & 0 & & \mu_{\rm m}^{*}
    \end{pmatrix},\notag\\
	\mathbb{J}_{\rm m}=\rho L_{\rm c}^2
	&\begin{pmatrix}
	\kappa_\gamma + \gamma_{1} & \kappa_\gamma - \gamma_{1} & \star & \dots & 0\\ 
	\kappa_\gamma - \gamma_{1} & \kappa_\gamma + \gamma_{1} & \star & \dots & 0\\ 
	\star & \star & \star & \dots & 0\\
	\vdots & \vdots & \vdots & \ddots &\\ 
	0 & 0 & 0 & & \gamma^{*}_{1}\\ 
	\end{pmatrix},
	&\mathbb{T}_{\rm e}=\rho L_{\rm c}^2
	&\begin{pmatrix}
	\kappabar_{\gamma} + \gammabar_{1} & \kappabar_{\gamma} - \gammabar_{1} & \star & \dots	& 0\\ 
	\kappabar_{\gamma} - \gammabar_{1} &   \kappabar_{\gamma} + \gammabar_{1} & \star & \dots & 0\\ 
	\star & \star & \star & \dots & 0\\
	\vdots & \vdots & \vdots & \ddots &\\ 
	0 & 0 & 0 & & \gammabar^{*}_{1}
	\end{pmatrix},\notag\\
    \mathbb{L}_{\rm s}=L_{\rm c}^2
    &\matr{
    \star                   & \star                   & \star                   & \multicolumn{2}{c}{\dots} & 0     \\
    \star                   & \star                   & \star                   & \multicolumn{2}{c}{\dots} & 0     \\
    \star                   & \star                   & \star                   & \multicolumn{2}{c}{\dots} & 0     \\
    \multirow{2}{*}{\vdots} & \multirow{2}{*}{\vdots} & \multirow{2}{*}{\vdots} & \alpha_1    & 0           & 0     \\
                            &                         &                         & 0           & \alpha_1    & 0     \\
    0                       & 0                       & 0                       & 0           & 0           & \star
    },
    &\mathbb{M}_{\rm s}=\rho \. L_{\rm c}^4
    &\matr{
    \star                   & \star                   & \star                   & \multicolumn{2}{c}{\dots} & 0     \\
    \star                   & \star                   & \star                   & \multicolumn{2}{c}{\dots} & 0     \\
    \star                   & \star                   & \star                   & \multicolumn{2}{c}{\dots} & 0     \\
    \multirow{2}{*}{\vdots} & \multirow{2}{*}{\vdots} & \multirow{2}{*}{\vdots} & \beta_1    & 0           & 0     \\
                            &                         &                         & 0           & \beta_1    & 0     \\
    0                       & 0                       & 0                       & 0           & 0           & \star
    },
    \label{eq:micro_ine_1}\\
	\noalign{\centering$\displaystyle\mathbb{C}_{\rm c}= 
    \begin{pmatrix}
    \star & 0 & 0\\ 
	0 & \star & 0\\ 
	0 & 0 & 4\.\mu_{\rm c}
    \end{pmatrix},\qquad\qquad
	\mathbb{J}_{\rm c}=\rho L_{\rm c}^2
	\begin{pmatrix}
	\star & 0 & 0\\ 
	0 & \star & 0\\ 
	0 & 0 & 4\.\gamma_{2}
	\end{pmatrix},\qquad\qquad
	\mathbb{T}_{\rm c}=\rho L_{\rm c}^2
	\begin{pmatrix}
    \star & 0 & 0\\ 
	0 & \star & 0\\ 
	0 & 0 & 4\.\gammabar_{2}
	\end{pmatrix},$\\
    $\mathbb{L}_{\rm a}=
    L_{\rm c}^2
    \begin{pmatrix}
    4\.\alpha_2 & 0 & 0\\ 
	0 & 4\.\alpha_2 & 0\\ 
	0 & 0 & \star
    \end{pmatrix},\qquad\qquad
    \mathbb{M}_{\rm a}=\rho \. L_{\rm c}^4
    \begin{pmatrix}
    4\.\beta_2 & 0 & 0\\ 
	0 & 4\.\beta_2 & 0\\ 
	0 & 0 & \star
    \end{pmatrix}.$}\notag
\end{align}
It is emphasized that the matrix representation of $\mathbb{L}_{\rm s}$, $\mathbb{L}_{\rm a}$, $\mathbb{M}_{\rm s}$, and $\mathbb{M}_{\rm a}$ differs from the others since the non zero terms in $(\Curl\.P)_{ij}$ and in $(\Curl\.\dot{P})_{ij}$ are the ``out of plane curvatures'' related to the indexes $i=1,2$ and $j=3$.
To better explain this fact, we consider the shape of $\Curl m$ given by
\begin{align}
    P &= 
    \begin{pmatrix}
    \bullet & \bullet & 0 \\ 
	\bullet & \bullet & 0 \\ 
	0 & 0 & 0
    \end{pmatrix},
    &\Curl\.P &=
    \begin{pmatrix}
    0 & 0 & \bullet\\ 
	0 & 0 & \bullet\\ 
	0 & 0 & 0
    \end{pmatrix},
    \label{eq:curl_shape}
    \\
    m =
    \left(\mathbb{L}_{\rm s} + \mathbb{L}_{\rm a}\right)\Curl\.P &= 
    \begin{pmatrix}
    0 & 0 & \bullet\\ 
	0 & 0 & \bullet\\ 
	\bullet & \bullet & 0
    \end{pmatrix},
    &\Curl\.m &=
    \begin{pmatrix}
    \bullet & \bullet & 0\\ 
	\bullet & \bullet & 0\\ 
	0 & 0 & \bullet
    \end{pmatrix}.
    \nonumber
\end{align}
Note that the structure of $\Curl\mhat=\Curl\left((\mathbb{M}_{\rm s}+\mathbb{M}_{\rm a})\Curl \ddot P\right)$ is identical.

As it can be seen from equation \eqref{eq:curl_shape}$_{4}$, even if the sym/skew decomposition is enforced in the constitutive law (we recall that $\mathbb{L}_{\rm s}$, $\mathbb{L}_{\rm a}$, $\mathbb{M}_{\rm s}$, and $\mathbb{M}_{\rm a}$ have minor symmetries, see Appendix~\ref{app:SymSkewCoup}), it is not guaranteed that the term $\Curl m$ appearing in the equilibrium equations \eqref{eq:equiMic} will retain only in-plane components for the most general constitutive tensor belonging to the tetragonal symmetry class.
In order to avoid this out of plane contribution, we must impose one of the following conditions
\begin{equation}
    \alpha_2 =\alpha_1\quad\wedge\quad\beta_2 =\beta_1
    \qquad\text{or}\qquad P_{22}=0\,.\label{eq:in_plane_constraint}
\end{equation}
Since the condition \eqref{eq:in_plane_constraint}$_{2}$ is not desirable because it prevents the presence of pressure waves, the condition \eqref{eq:in_plane_constraint}$_{1}$ will be adopted.
This choice, while reducing the number of parameters in the model, does not reduce the number of the independent ones for $\Curl P$ and $\Curl\dot P$ as shown in Appendix~\ref{app:SymSkewCoup}.

Furthermore, we can always compute the parameters\footnote{We write \enquote{$m$} for \enquote{$micro$} and \enquote{$M$} for \enquote{$macro$} for the corresponding elastic parameters to shorten the following expressions.} of the meso-scale depending on the micro-parameters and a new set of macro-parameters \cite{dagostino_effective_2020,neff_identification_2020,rizzi_exploring_2021}
\begin{align}
    \mathbb{C}_{\rm e}=	\mathbb{C}_{\rm micro}\,(\mathbb{C}_{\rm micro}-\mathbb{C}_{\rm macro})\inv\mathbb{C}_{\rm macro}\,,
    \label{eq:homo_formulae}
\end{align}
where $\mathbb{C}_{\rm macro}$ has the same structure of $\mathbb{C}_{\rm micro}$ and $\mathbb{C}_{\rm e}$.
For the tetragonal class of symmetry in 2D, the relation \eqref{eq:homo_formulae} particularize to
\begin{equation}
    	\mu_{\rm e}=\dfrac{\mu_{\rm m}\,\mu_{\rm M}}{\mu_{\rm m}-\mu_{\rm M}}\,,\qquad	
    	\kappa_{\rm e}=\dfrac{\kappa_{\rm m}\,\kappa_{\rm M}}{\kappa_{\rm m}-\kappa_{\rm M}}\,,\qquad
    	\mu_{\rm e}^*=\dfrac{\mu_{\rm m}^*\,\mu_{\rm M}^*}{\mu_{\rm m}^*-\mu_{\rm M}^*}\,,
\end{equation}
In the limit of infinitesimal small unit-cells or rather of an indefinitely large body, these macro-parameters are obtained as the homogenization to a classical Cauchy material \cite{dagostino_effective_2020,neff_identification_2020} and can directly be obtained by the slope of the acoustic curves at the origin, cf.\ Section~\ref{sec:fitting}.
Throughout this paper all material parameters introduced will be positive to guarantee the positive definiteness of their corresponding tensors.
%
%
%
%
%
\section{Dispersion curves}\label{sec:disperionCurbes}

We assume a plane strain\footnote{The components of $u$ and $P$ depends only on $\{x_1,x_2\}$ and $u_3=P_{13}=P_{31}=P_{23}=P_{32}=P_{33}=0$.} time harmonic ansatz for the displacement $u$ and the micro-distortion tensor $P$
\begin{equation}
    v_j=\Psi_j \. e^{i\left( k_1\.x_1 + k_2\.x_2 - \omega \. t \right)},
    \label{eq:time_harm_ansatz}
\end{equation}
where $v_j$ represents the generic component of $u$ or $P$, $\Psi_j$ is a scalar amplitude, $(k_1,k_2)^T=k\,(\sin \phi,\cos \phi)^T$ are the wavevector components with $\phi$ as the angle giving the direction of propagation, $k$ the wavevector length, and $\omega$ is the frequency.

Substituting the ansatz \eqref{eq:time_harm_ansatz} in the equilibrium equations \eqref{eq:equiMic}, we obtain the homogeneous algebraic linear system
\begin{equation}
    A \. \Psi =0
    \label{eq:eigen_prob}
\end{equation}
where $A=A(\omega,k,\phi) \in \mathbb{C}^{6\times6}$ is the acoustic tensor which depends on the frequency $\omega$, the wave vector length $k$, the angle of propagation $\phi$, all the constitutive parameters in equation \eqref{eq:micro_ine_1}, and $\Psi \in \mathbb{R}^{6}$ is the vector of amplitudes.\footnote{It is remarked that this is only possible since the condition \eqref{eq:in_plane_constraint} is satisfied.}
The non-trivial solutions of the system \eqref{eq:eigen_prob} are obtained when $A$ is singular, i.e.\ when $\det A=0$, which provides relations between $k=k(\omega,\phi)$ (or $\omega=\omega(k,\phi)$), the so-called dispersion relations.

The acoustic tensor can be written as
\begin{align}
A=
\left(
\begin{array}{ccc}
B_1 & C_1
 \\
C_2 & B_2
\end{array}
\right),
\end{align}
where $(B_1,B_2,C_1,C_2)\in \mathbb{C}^{3\times3}$.

Since $\det A = \det(B_1)\.\det(B_2-C_2\.B_1^{-1}\.C_1)$ \cite{Bernstein2009}, it is clear that the determinant of $A$ becomes the product of the two independent factors $\det B_1$ and $\det B_2$ if either $C_1=0$ or $C_2=0$.
If the reference system is chosen to be aligned with the direction of the wave vector (i.e.\ $\phi=\theta$, with $\theta$ the angel of rotation of the reference system), the sub-matrices $C_1$ and $C_2$ expressions are
\begin{align}
C_1
&=
\frac{\sin(4\theta)}{2}
\left(
\begin{array}{ccc}
 k^2
 \left(q_{\rm e}-L_{\rm c}^2 \rho \. \overline{q}_\gamma \omega ^2\right)
 &
 i \. k \. q_{\rm e}
 &
 i \. k \. q_{\rm e}
 \\
 -i \. k \. q_{\rm e}
 &
 - \rho \. L_{\rm c}^2 \. q_\gamma \. \omega^2 + q_{\rm e} + q_{\rm m}
 &
 - \rho \. L_{\rm c}^2 \. q_\gamma \. \omega^2 + q_{\rm e} + q_{\rm m}
 \\
 i \. k \. q_{\rm e}
 &
  \rho \. L_{\rm c}^2 \. q_\gamma \. \omega ^2 - q_{\rm e} - q_{\rm m}
 &
  \rho \. L_{\rm c}^2 \. q_\gamma \. \omega ^2 - q_{\rm e} - q_{\rm m} \\
\end{array}
\right),
\label{eq:acoustic_tens_extra_diag}
\\*
C_2
&=
\frac{\sin(4\theta)}{2}
\left(
\begin{array}{ccc}
 k^2 \left(q_{\rm e}-L_{\rm c}^2 \rho  \overline{q}_\gamma \omega ^2\right)
 &
 i \. k \. q_{\rm e}
 &
 -i \. k \. q_{\rm e}
 \\
 -i \. k \. q_{\rm e}
 &
 -\rho \. L_{\rm c}^2 \. q_\gamma \. \omega^2 + q_{\rm e} + q_{\rm m}
 &
  \rho \. L_{\rm c}^2 \. q_\gamma \. \omega^2 - q_{\rm e} - q_{\rm m}
 \\
 -i \. k \. q_{\rm e}
 &
 -\rho \. L_{\rm c}^2 \. q_\gamma \. \omega^2 + q_{\rm e} + q_{\rm m}
 &
  \rho \. L_{\rm c}^2 \. q_\gamma \. \omega^2 - q_{\rm e} - q_{\rm m}
 \\
\end{array}
\right),
\notag
\end{align}
where $q_{\rm e}=\mu_{\rm e}-\mu_{\rm e}^{*}\,,\;q_{\rm m}=\mu_{\rm m}-\mu_{\rm m}^{*}\,,\;q_\gamma=\gamma_1-\gamma_1^{*}\;\text{and}\;\overline{q}_\gamma=\gammabar_1-\gammabar_1^{*}$.
It is highlighted that, since the constitutive law for the curvature terms Curl$P$ and Curl$\dot P$ depends on just one parameter, they are isotropic, and this makes the matrices $C_1$ and $C_2$ independent with respect $\alpha_i$ and $\beta_i$ while the expressions of the latters in $B_1$ and $B_2$ are not affected by the rotation of the reference system.

From equation \eqref{eq:acoustic_tens_extra_diag} it is possible to deduce that the condition for which $C_1=C_2=0$ (one of the two would be already enough) is $\theta=(\pi \. n)/4$ with $n\in\N$.
This means that, when the reference system is aligned with the direction of wave propagation, and both are aligned with a symmetry axes of the material, the determinant $\det A= \det(B_1)\.\det(B_2)$ is the product of two independent factors.

\textit{These two independent factors can be associated with pure-pressure waves $\det(B_1)$ and pure-shear waves $\det(B_2)$.}

In particular, this allows us to reduce the order of the dispersion polynomial by
\begin{equation}
    \det A=p(k^2,k^4,k^6,k^8,\omega^2,\omega^4,\omega^6,\omega^8,\omega^{10},\omega^{12})=p_1(k^2,k^4,\omega^2,\omega^4,\omega^6)\.p_2(k^2,k^4,\omega^2,\omega^4,\omega^6)\,,
\end{equation}
where $p_1$ and $p_2$ are of the form
\begin{equation}
    p_i=
    c_0\.k^2+c_1\.k^4-(c_2+c_3\.k^2+c_4\.k^4)\,\omega^2+(c_5+c_6\.k^2+c_7\.k^4)\,\omega^4-(c_8+c_9\,k^2+c_{10}\.k^4)\,\omega^6,
    \label{eq:dispRelationAll}
\end{equation}
easing the calculation of the roots of these polynomials significantly, i.e.\ the expressions of the dispersion curves. It is noted that we can always consider $k(\omega)$, instead of $\omega(k)$, for shorter analytical expression as we must only solve a quadratic equation (in $k^2$) and not a third order polynomial (in $\omega^2$) but use $\omega(k)$ for easier plotting and its natural split in three distinct expressions for each dispersion curve.
%
%
%
%
%
\section{New considerations on the relaxed micromorphic parameters}\label{sec:parameters}

In this section, we draw some useful considerations about the consistency of the relaxed micromporpic model with respect to a change of unit cell's size and of the material properties of the base material. The model's consistency is checked against a standard Bloch-Floquet analysis of the wave propagation performed using the unit cell described in Section~\ref{sec:unitCell} with built in periodic Bloch-Floquet boundary conditions from \comsol.

The following two connections between the properties of the unit cell and the behaviour of the dispersion curves can be drawn:
\begin{itemize}
    \item The dispersion curves scale proportionally in $\omega$ with respect to the speed of the wave of the bulk material composing the unit cell;
    \item The dispersion curves scale inversely in both $\omega$ and $k$ with respect to the size of the unit cell.
\end{itemize}

Both results are useful to avoid repeating the time-consuming fitting procedure when changing the size of the cell and the base material's properties while keeping the unit cell's geometry unchanged.

\subsection{Consistency of the relaxed micromorphic model with respect to a change in the unit cell's bulk material properties}
The dispersive properties of a microstructured isotropic Cauchy material depend exclusively and linearly on the waves speeds of the bulk material once its geometry is fixed. Indeed, as it is well known, the dispersion relatives can be always written as $\omega=k\.c_i$, $i=\{p,s\}$, where  $c_{\rm p}$ and $c_{\rm s}$ are the pressure and shear wave speeds, respectively, defined as
\begin{equation}
    c_{\rm p} \coloneqq
    \sqrt{\frac{\kappa_{\rm M}+\mu_{\rm M}}{\rho}}\,,
    \qquad
    c_{\rm s} \coloneqq
    \sqrt{\frac{\mu_{\rm M}}{\rho}}\,,
    \qquad
    k=\frac{\omega}{c_i}\,,
    \qquad
    i=\{p,s\}\,.
    \label{eq:speedWaves}
\end{equation}
This implies that by scaling the elastic coefficients by a constant $a$ and the density by another constant $b$ we have
\begin{equation}
    \widetilde{c}_{\rm p} \coloneqq
    c_{\rm p}
    \.
    \sqrt{\frac{a}{b}}
    \,,
    \qquad
    \widetilde{c}_{\rm s} \coloneqq
    c_{\rm s}
    \.
    \sqrt{\frac{a}{b}}
    \,,
    \qquad
    k=\sqrt{\frac{b}{a}}\.
    \frac{\omega}{c_i}\,,
    \qquad
    i=\{p,s\}\,.
    \label{eq:speedWaves2}
\end{equation}
Therefore, the response of an effective model should also
change accordingly.
Thus, we observe that by multiplying all the relaxed micromorphic elastic coefficients ($\mathbb{C}_{\rm e},\mathbb{C_{\rm c}},\mathbb{C}_{\rm micro},\mathbb{L}_{\rm s},\mathbb{L}_{\rm a}$) by a constant $a$ and the apparent density $\rho$ by another constant $b$ we can rewrite equation \eqref{eq:dispRelationAll} as\footnote{Scaling the density $\rho$ automatically scale all the micro-inertia terms too since all of them are already proportional to it.}
\begin{align}
    a^3(c_0\.k^2+c_1\.k^4)
    -a^2\.b\.(c_2+c_3\.k^2+c_4\.k^4)\,\omega^2
    +a\.b^2\.(c_5+c_6\.k^2+c_7\.k^4)\,\omega^4 
    \hspace{1.5cm}
    \label{eq:DisperionsScaleSpeed}
    \\
    -b^3\.(c_8+c_9\,k^2+c_{10}\.k^4)\,\omega^6 &=0\,.
    \nonumber
\end{align}
By collecting $a^3$ we obtain
\begin{align}
    a^3
    \left[
    (c_0\.k^2+c_1\.k^4)
    -(c_2+c_3\.k^2+c_4\.k^4)
    \left(\omega \sqrt{\frac{b}{a}}\right)^2
    +(c_5+c_6\.k^2+c_7\.k^4)
    \left(\omega \sqrt{\frac{b}{a}}\right)^4 
    \hspace{1.5cm}
    \right.
    \label{eq:DisperionsScaleSpeed2}
    \\
    \left.
    -\.(c_8+c_9\,k^2+c_{10}\.k^4)
    \left(\omega \sqrt{\frac{b}{a}}\right)^6 
    \right]
    &=0\,,
    \nonumber
\end{align}
from which wa can introduce a scaled frequency $\widetilde\omega=\omega\.\sqrt{\frac{b}{a}}$. Increasing the stiffness ($a$) or decreasing the density ($b$) of the base material will cause an overall shifting of the dispersion curves towards lower frequencies.

This is consistent with what is observed looking at the dispersion properties of a microstructured material's unit cell obtained via a Bloch-Floquet analysis.
In the case for which $a=b$, the roots of equation \eqref{eq:DisperionsScaleSpeed2} do not change at all.
Thanks to this identification, we can now easily change the material constituting the unit cell (without changing the geometry) by scaling the material parameters accordingly without repeating the whole fitting process. In particular, all the cut-offs and asymptotes will be scaled by a quantity $\sqrt{\frac{b}{a}}$.

We explicitly remark again that scaling the macroscopic apparent density $\rho$ of the unit cell by a factor $b>0$ will change the frequency $\omega$ by the factor $\frac{1}{\sqrt b}$, i.e.\ the frequency is inversely proportional to the square root of the density of the unit cell. The wavenumber $k$ is invariant under changing the macroscopic apparent density since the periodicity of the unit cell remains unaltered.

\subsection{Consistency of the relaxed micromorphic model with respect to a change in the unit cell's size}
While keeping the geometry and the material unaltered, the dispersion properties of a microstructured isotropic Cauchy material are inversely proportional to the size of its unit cell, meaning that halving the size of the unit cell will double the frequency response for each value of the length $k$ of the wavevector, which also changes with the same inverse proportionality since it represents the spatial periodicity of the structure. This can be easily retrieved by performing standard Bloch-Floquet analysis.

In order to obtain this behaviour with the relaxed micromorphic model, we must scale the elastic curvature tensors ($\mathbb{L}_{\rm s},\mathbb{L}_{\rm a}$) and all the micro-inertia tensors ($\mathbb{J}_{\rm m},\mathbb{J}_{\rm c}$,$\mathbb{T}_{\rm e},\mathbb{T}_{\rm c}$) by the square of the size of the unit cell, and the micro-inertia curvature tensors ($\mathbb{M}_{\rm s},\mathbb{M}_{\rm a}$) by the fourth power of the size of the unit cell.

To prove this, we consider a change in the size of the unit cell by some arbitrary factor $t>0$, which requires the scaling of the characteristic length $L_{\rm c}$ (which is now considered to be equal to the size of the unit cell) by the same factor $t$. Assuming that all the other material parameters used remain constant, in equation \eqref{eq:dispRelationAll} we substitute $L_{\rm c}\to t\.L_{\rm c}$ (see the coefficients in Appendix~\ref{app:coefficients}-\ref{app:coefficients3})
\begin{align}
    &c_0\.k^2+t^2\.c_1\.k^4-c_2\.\omega^2+t^2\.c_3\.k^2\.\omega^2+t^4\.c_4\.k^4\.\omega^2+t^2\.c_5\.\omega^4\label{eq:DisperionsScaleLc}\\
    &\;+t^4\.c_6\.k^2\.\omega^4+t^6\.c_7\.k^4\.\omega^4-t^4\.c_8\.\omega^6+t^6\.c_9\.k^2\.\omega^6+t^8\.c_{10}\.k^4\.\omega^6=0\,.\notag
\end{align}
We can now collect $\frac{1}{t^2}$ in \eqref{eq:DisperionsScaleLc} arriving at
\begin{align}
    \frac{1}{t^2}\bigl[&c_0( t\.k)^2
    +c_1\.(t\.k)^4
    -c_2\.(t\.\omega)^2
    +c_3\.(t\.k)^2(t\.\omega)^2
    +c_4\.(t\.k)^4(t\.\omega)^2
    +c_5\.(t\.\omega)^4\label{eq:DisperionsScaleLc2}\\
    &+c_6\.(t\.k)^2(t\.\omega)^4
    +c_7\.(t\.k)^4(t\.\omega)^4
    -c_8\.(t\.\omega)^6
    +c_9\.(t\.k)^2(t\.\omega)^6
    +c_{10}\.(t\.k)^4(t\.\omega)^6\bigr]=0\,.\notag
\end{align}

It is clear to see from the comparison between equation \eqref{eq:DisperionsScaleLc} and equation \eqref{eq:DisperionsScaleLc2} that their roots are simply linearly scaled with respect the factor $t$.
This result enables the choice of having $L_{\rm c}$ equal to the size of the unit cell and thus allows us to change the latter in the microstructured material considered without needing to repeat the whole fitting procedure.

In particular, scaling the size $L_{\rm c}$ of the unit cell by the factor $t>0$ will change the frequency $\omega$ and wavenumber $k$ by the factor $\frac{1}{t}$, i.e.\ the frequency and wavenumber are reverse proportional to the size of the unit cell. This simple observation allows to perform the fitting procedure for the relaxed micromorphic model only once for each geometry of the unit cell: changing the size of the unit cell will result in an automatic fitting when suitably rescaling $\omega$ and $k$ where none of the material parameters (except $L_{\rm c}$) must be changed. Note that the slopes at the origin of the dispersion curves do not change when changing the size of the unit cell while keeping the geometry fixed. 
%
%
%
%
%
\subsection{Relaxed micromorphic cut-offs}\label{sec:cut-offs}

The cut-offs of the dispersion curves play an important role in fitting the material parameters of the relaxed micromorphic model \cite{neff_identification_2020,dagostino_effective_2020,rizzi_exploring_2021}. For the convenience of the reader, we show the calculations of the analytic expressions again. In the case $k=0$, the dispersion relation \eqref{eq:dispRelationAll} simplifies into
\begin{gather}
    -c_2\,\omega^2+c_5\,\omega^4-c_8\,\omega^6=0
    \quad
    \iff
    \quad
    \omega^2\left(\omega^4-\frac{c_5}{c_8}\.\omega^2+\frac{c_2}{c_8}\right)=0\notag
    \\
    \quad
    \iff
    \quad
    \omega_1^2=0 \,,
    \quad
    \omega^2_{2,3}=\dfrac{c_5\pm\sqrt{c_5^2-4\.c_2c_8}}{2\.c_8}\,.
    \label{eq:cutoffsStart}
\end{gather}
The coefficients $c_2,c_5,c_8$ depend on the elastic parameters $\mu_{\rm m},\kappa_{\rm m},\mu_{\rm m}^*,\mu_{\rm c},\mu_{\rm e},\kappa_{\rm e},\mu_{\rm e}^*$, the micro-inertia parameters $\kappa_\gamma, \gamma_1, \gamma_1^*, \gamma_2$, the macroscopic apparent density $\rho$, and characteristic length $L_{\rm c}$ but are independent of the parameters for $\Curl P$, $\grad u$, and $\Curl\dot P$, cf.\ Appendix~\ref{app:coefficients}.

Equations \eqref{eq:cutoffsStart} can be simplified as in Table~\ref{tab:cut-offs} with
\begin{align}
    \omega_{\rm r}=\sqrt{\frac{\mu_{\rm c}}{\rho\.L_{\rm c}^2\.\gamma_2}}\,,
    \qquad
    \omega_{\rm s}=\sqrt{\frac{\mu_{\rm e}+\mu_{\rm m}}{\rho\.L_{\rm c}^2\.\gamma_1}}\,,
    \qquad
    \omega_{\rm ss}=\sqrt{\frac{\mu_{\rm e}^*+\mu_{\rm m}^*}{\rho\.L_{\rm c}^2\.\gamma_1^*}}\,,
    \qquad
    \omega_{\rm p}=\sqrt{\frac{\kappa_{\rm e}+\kappa_{\rm m}}{\rho\.L_{\rm c}^2\.\kappa_\gamma}}\,.
    \label{eq:cut-offs}
\end{align}
\begin{table}[h!]
\centering
\begin{tabular}{c|c|c}
 &
  $0^\circ$ &
  $45^\circ$ \\ \hline
shear &
  \begin{tabular}[c]{@{}c@{}}
  $\omega_2=\omega_{\rm ss}$\\
  $\omega_3=\omega_{\rm r}$
  \end{tabular} &
  \begin{tabular}[c]{@{}c@{}}
  $\omega_2=\omega_{\rm s}$\\
  $\omega_3=\omega_{\rm r}$
  \end{tabular}
\end{tabular}
\qquad
\begin{tabular}{c|c|c}
 &
  $0^\circ$ &
  $45^\circ$ \\ \hline
pressure &
  \begin{tabular}[c]{@{}c@{}}
  $\omega_2=\omega_{\rm s}$\\
  $\omega_3=\omega_{\rm p}$
  \end{tabular} &
  \begin{tabular}[c]{@{}c@{}}
  $\omega_2=\omega_{\rm ss}$\\
  $\omega_3=\omega_{\rm p}$
  \end{tabular}
\end{tabular}
\caption{Cut-offs expressions for the pressure waves (left) and for the shear waves (right).}
\label{tab:cut-offs}
\end{table}

We recognize that the expressions $\omega_{\rm s}$ and $\omega_{\rm ss}$ change from pressure to shear and shear to pressure, respectively, when going from 0 to 45 degrees of incidence.
Since the dispersion curves of the unit cell have two cut-offs that coincide, we chose them to be $\omega_{\rm s}$ and $\omega_{\rm ss}$.
Therefore, we introduce the following relation
\begin{equation}
    \omega_{\rm s}=\omega_{\rm ss}\qquad\iff\qquad\frac{\mu_{\rm e}+\mu_{\rm m}}{\gamma_1}=\frac{\mu_{\rm e}^*+\mu_{\rm m}^*}{\gamma_1^*}\,.
\end{equation}
The values of theses cut-offs have been fixed according to \comsol simulations as
\begin{align}
    \omega_{\rm r}=554.61~\si{\Hz}\,,
    \qquad
    \omega_{\rm s}=\omega_{\rm ss}=2011.83~\si{\Hz}\,,
    \qquad
    \omega_{\rm p}=2048.19~\si{\Hz}\,,
    \label{eq:cut-offs_values}
\end{align}
and the values of last points from \comsol are used to fix the asymptotes, cf.\ Table \ref{tab:asymptotesValues}.
\renewcommand{\arraystretch}{1.5}
\begin{table}[ht!]
    \centering
    \begin{tabular}{c|c|c}
         & $0^\circ$ & $45^\circ$ \\ \hline
         & 340.42~\si{\Hz}& 606.31~\si{\Hz}   \\
shear    & 637.79~\si{\Hz}  & 689.21~\si{\Hz}   \\
         & 2078.39~\si{\Hz}  & 2375.78~\si{\Hz}
    \end{tabular}
    \qquad
    \begin{tabular}{c|c|c}
         & $0^\circ$ & $45^\circ$ \\ \hline
         & 581.95~\si{\Hz}  & 689.20~\si{\Hz}   \\
pressure & 2040.81~\si{\Hz}  & 1883.97~\si{\Hz}   \\
         & 2156.21~\si{\Hz}  & 2376.45~\si{\Hz} 
    \end{tabular}
    \caption{Numerical values of the asymptotes via Bloch-Floquet analysis using \comsol.}\label{tab:asymptotesValues}
\end{table}
\renewcommand{\arraystretch}{1}
%
%
%
%
%
\section{Fitting of the relaxed micromorphic parameters: the particular case of vanishing curvature (without $\Curl P$ and $\Curl\dot P$)}\label{sec:fitting1}

In the numerical applications considered in this work, we start without considering the tensors $\mathbb{L}_{\rm s}, \mathbb{L}_{\rm c}$ and $\mathbb{M}_{\rm s}, \mathbb{M}_{\rm c}$, i.e.\ we neglect the effect of $\Curl P$ and $\Curl\dot P$ on the dynamic regime. This fundamentally changes the shape of the analytical expression \eqref{eq:dispRelationAll} resulting in a new polynomial $p(k^2,\omega^2,\omega^4,\omega^6)$ with reduced order of $k$. Assuming that the wavenumber is always positive, we only have one single expression $k(\omega)$ describing all three dispersion curves
\begin{gather}
    c_0\.k^2-(c_2+c_3^*\.k^2)\.\omega^2+(c_5+c_6^*\.k^2)\.\omega^4-(c_8+c_9^*\,k^2)\.\omega^6=0
    \,,\quad
    \label{eq:dispRelation}
    k=\omega\.\sqrt{\frac{c_2-c_5\.\omega^2+c_8\.\omega^4}{c_0-c_3^*\.\omega^2+c_6^*\.\omega^4-c_9^*\.\omega^6}}\,.
\end{gather}
The coefficients $c_0,c_2,c_3^*,c_5,c_6^*,c_8,c_9^*$ used\footnote{The notation $c_i^*$ throughout the following sections indicates that enhancing the model, i.e.\ reintroducing the neglected tensors for $\Curl P$ and $\Curl\dot P$, changes $c_i^*$ by the addition of new terms. In contrast, the notation $c_i$ without an asterisk indicates that the coefficient is complete in the sense that it remains unchanged when including $\Curl P$ and $\Curl\dot P$ as well.} have a different expression at $0^\circ$ and $45^\circ$ angle of incidence as well as for pressure and shear waves and are included in the Appendix~\ref{app:coefficients}.
The structure of $k(\omega)$ in equation \eqref{eq:dispRelation}$_{2}$ can be also reported as a non-linear classic dispersion relation $k=\frac{\omega}{c_{\text{p/s}}(\omega)}$ with a frequency dependent group velocity $c_{\text{p/s}}$.

\subsection{Asymptotes}\label{sec:asymptotes1}
Instead of fitting dispersion curves pointwise, we focus on using the analytical expression of the cut-offs ($k=0$) and of the asymptotes ($k\to\infty$). The explicit expression of the cut-offs is already discussed in Section~\ref{sec:cut-offs}. We use a similar approach to calculate the asymptotes as well by considering the limit $k\to\infty$ where only the terms with the highest order of $k$ are important. Thus, we arrive at
\begin{align}
    c_0-c_3^*\.\omega^2+c_6^*\.\omega^4-c_9^*\.\omega^6=0\qquad\iff\qquad\omega^6-\frac{c_6^*}{c_9^*}\.\omega^4+\frac{c_3^*}{c_9^*}\.\omega^2-\frac{c_0}{c_9}=0\,.\label{eq:asymptotes}
\end{align}
In contrast to the analytical expression of the cut-offs, we are not able to simplify the asymptotes' expression in a feasible way. This is mainly due to the fact that we must solve a third order polynomial while we only had two non-zero cut-offs each before.

Since the dispersion curves of the unit cell obtained via Bloch-Floquet analysis are by nature periodic, the limit for $k\to\infty$ is \textit{per se} meaningless when considering the Bloch-Floquet approach. The value for $k=1/L_{\rm c}$ (with $L_{\rm c}$ the size of the unit cell) is the periodicity limit.
On the other hand, this limit has of course meaning for a continuum model like the relaxed micromorphic model.
To reconcile these two limits in the fitting procedure, we will impose that the limit $k\to\infty$ for our continuum model will coincide with the periodicity limit of the Bloch-Floquet curves $k=1/L_{\rm c}$.
This strategy allows us to preserve the width of the band-gap, cf.\ Figure \ref{fig:fittingProcedure}.
\newcommand{\sample}{20}
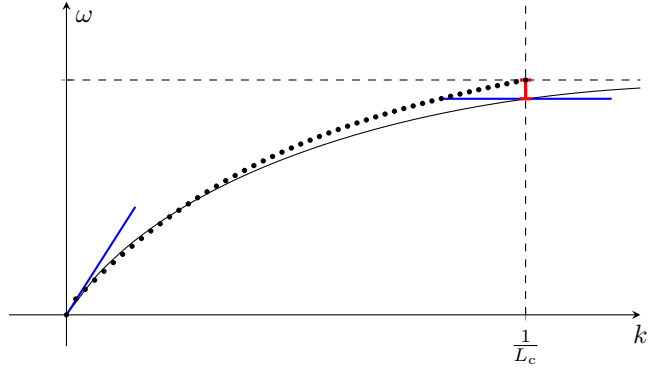
\begin{figure}[ht!]
    \centering
    \begin{tikzpicture}
        \begin{axis}[
        axis x line=middle,axis y line=middle,
        x label style={at={(current axis.right of origin)},anchor=north, below},
        xlabel=$k$, ylabel=$\omega$,
        xmin=-1, xmax=10,
        ymin=-0.4, ymax=4,
        width=0.6\linewidth,
        height=0.6*0.618*\linewidth,
        ytick={3},
        xtick={8},
        xticklabels={$\frac{1}{L_{\rm c}}$},
        yticklabels={,}
        ]
        \addplot[only marks, mark=*, mark options={solid,black,scale=0.4}][domain=0:8, samples=50]{ln(x+1)+0.15*x+(1.8-ln(9))/64*x^2+0.05*sin(270*x^(1/3))};
        \addplot[black, smooth][domain=0:10, samples=\sample]{ln(x+1)+0.15*x-0.01*x^2};
        \addplot[black, dashed] coordinates {(0, 3) (10, 3)};
        \addplot[black, dashed] coordinates {(8, 0) (8, 4)};
        \addplot[blue, smooth, thick][domain=0:1.2,samples=\sample]{1.15*x};
        \addplot[blue, smooth, thick] coordinates {(6.5, 2.76) (9.5, 2.76)};
        \addplot[red, smooth, very thick] coordinates {(8, 2.76) (8, 3)};
        \addplot[red, smooth, very thick] coordinates {(7.9, 2.76) (8.1, 2.76)};
        \addplot[red, smooth, very thick] coordinates {(7.9, 3) (8.1, 3)};
        \end{axis}
    \end{tikzpicture}
    \caption{The fitting of the dispersion curves (black line) includes information about the slope at zero and the asymptotes (blue lines) while the remaining shape comes automatically. The Block-Floqeut analyis used in \comsol\ can only evaluate up to $k=\frac{1}{L_{\rm c}}$ (black dots) which causes a slight gap between the dispersion curves at $k=\frac{1}{L_{\rm c}}$ and their corresponding asymptotes (red part).}
    \label{fig:fittingProcedure}
\end{figure}

\subsection{Fitting}\label{sec:fitting}
We start the fitting with the macroscopic apparent density $\rho$ and values of the macro parameters $\kappa_{\rm M},\mu_{\rm M},\mu_{\rm M}^*$, i.e.\ the material constants necessary for the classical homogenization of an infinite large micromorphic material. For an anisotropic Cauchy material, the speed of the acoustic waves is
\begin{align}
    c_{\rm p}&=\sqrt{\frac{\kappa_{\rm M}+\mu_{\rm M}}{\rho}}\,,&c_{\rm s}&=\sqrt{\frac{\mu_{\rm M}^*}{\rho}}\,,\\
    \cbar_{\rm p}&=\sqrt{\frac{\kappa_{\rm M}+\mu_{\rm M}^*}{\rho}}\,,&\cbar_{\rm s}&=\sqrt{\frac{\mu_{\rm M}}{\rho}}\,,\notag
\end{align}
where $c_p,c_s$ are the speed of pressure and shear wave, respectively, for $0^\circ$ of incidence while $\cbar_p,\cbar_s$ describe and incidence angle of $45^\circ$. For the tetragonal class of symmetry we choose, it holds
\begin{equation}
    c_{\rm p}^2+c_{\rm s}^2=\frac{\kappa_{\rm M}+\mu_{\rm M}+\mu_{\rm M}^*}{\rho}=\cbar_{\rm p}^2+\cbar_{\rm s}^2,
\end{equation}
reducing the system of equations to just three independent quantities. For the relaxed micromorphic model we fit these macro parameters by the slope of the corresponding acoustic dispersion curves at $k=0$
\begin{equation}
    \mu_{\rm M}=\cbar_{\rm s}^2\.\rho\,,\qquad\mu_{\rm M}^*=c_{\rm s}^2\.\rho\,,\qquad\kappa_{\rm M}=(c_{\rm p}^2-\cbar_{\rm s}^2)\.\rho=(\cbar_{\rm p}^2-c_{\rm s}^2)\.\rho\,.
\end{equation}
The remaining unknown density $\rho$ is given by the material and geometry of the cell we used (see Figure~\ref{fig:ChartesGeometry}) and is directly computed as $\rho=\overline\rho\.\frac{A_{\rm total}-A_{\rm voids}}{A_{\rm total}}$ with $\overline\rho$ as the density of Polyethylene and $A_{\rm total}$, $A_{\rm voids}$ being the area of the whole cell and its voids, respectively. We list the numerical values in Table~\ref{tab:numericalValuesMacro2}.

\renewcommand{\arraystretch}{1.5}
\begin{table}[ht!]
    \centering
    \begin{tabular}{c|c|c|c|c}
        $L_{\rm c}$ & $\rho$ & $\kappa_{\rm M}$ & $\mu_{\rm M}$ & $\mu_{\rm M}^*$\\ \hline
        $[\si{\mm}]$ & $[\si[per-mode = symbol]{\kg\per\cubic\m}]$ & $[\si{\kPa}]$ & $[\si{\kPa}]$ & $[\si{\kPa}]$\\ \hline
        20 & 361.22 & 88.06 & 50.01 & 24.84
    \end{tabular}
    \caption{Numerical values of the macroscopic apparent density $\rho$ and the elastic macro-parameters $\kappa_{\rm M},\mu_{\rm M},\mu_{\rm M}^*$ for our unit cell when considering Polyethylene as base material.}
    \label{tab:numericalValuesMacro2}
\end{table}
\renewcommand{\arraystretch}{1}

As a second step, we use the analytical expressions for the cut-offs \eqref{eq:cut-offs} and calculate
\begin{align}
    \kappa_\gamma&=\frac{\kappa_{\rm e}+\kappa_{\rm m}}{\rho\.L_{\rm c}^2\.\omega_p^2}\,,
    &\gamma_1&=\frac{\mu_{\rm e}+\mu_{\rm m}}{\rho\.L_{\rm c}^2\.\omega_s^2}\,,\\
    \gamma_1^*&=\frac{\mu_{\rm e}^*+\mu_{\rm m}^*}{\rho\.L_{\rm c}^2\.\omega_{ss}^2}\,,
    &\gamma_2&=\frac{\mu_{\rm c}}{\rho\.L_{\rm c}^2\.\omega_r^2}\,.\notag
\end{align}
Thus, we can reduce the system of independent variables by the four inertia parameters $\kappa_\gamma,\gamma_1,\gamma_1^*,\gamma_2$. Note again that we implied $\omega_{\rm s}=\omega_{\rm ss}$, see Section \ref{sec:cut-offs}.

As the third step, we use the analytic expression of the asymptotes for zero and 45 degrees resulting in 12 expression in total which is more then the number of independent parameters remaining. Thus we fit the last 8 remaining parameters, namely the micro parameters $\kappa_{\rm m},\mu_{\rm m},\mu_{\rm m}^*,\mu_{\rm c}$ and inertia parameters $\kappabar_\gamma,\gammabar_1,\gammabar_1^*,\gammabar_2$, numerically by minimizing the square error of the analytical expressions \eqref{eq:asymptotes} and their corresponding numerical values from \comsol\ (for $k=\frac{1}{L_{\rm c}}$ for zero degree and $k=\frac{\sqrt2}{L_{\rm c}}$ for 45 degrees). Thereby, we utilize the fact that each group of asymptotes only depend on 4 independent parameters to speed up the calculations, as summarized in Table~\ref{tab:dependencyParameters}.

\renewcommand{\arraystretch}{1.5}
\begin{table}[ht!]
\centering
\begin{tabular}{l|ll}
    & pressure &  shear \\ \hline
     $0^\circ$ & $\kappa_{\rm m},\kappabar_\gamma,\mu_{\rm m},\gammabar_1$ & $\mu_{\rm c},\gammabar_2,\mu_{\rm m}^*,\gammabar_1^*$\\
    $45^\circ$ & $\kappa_{\rm m},\kappabar_{\gamma},\mu_{\rm m}^*,\gammabar_1^*$ & $\mu_{\rm c},\gammabar_2,\mu_{\rm m},\gammabar_1$          
\end{tabular}
\caption{Dependence of the asymptotes of the dispersion curves on the free material parameters as function of the direction of propagation ($0^\circ$/$45^\circ$) and type of wave (shear/pressure).}
\label{tab:dependencyParameters}
\end{table}
\renewcommand{\arraystretch}{1}

We calculate these values with \mathematica\ using the \textit{NMinimize}-algorithm with the inbuilt method \textit{RandomSearch} running in a loop for multiple times. Here, we must start with reasonable initial values for all parameters involved. Note that it is always possible to imply $\mathbb{C}_{\rm micro}=2\.\mathbb{C}_{\rm macro}$ as a conservative first guess implying 
\begin{equation}
    \mathbb{C}_{\rm e}=\mathbb{C}_{\rm micro}\,(\mathbb{C}_{\rm micro}-\mathbb{C}_{\rm macro})\inv\mathbb{C}_{\rm macro}=2\,\mathbb{C}_{\rm macro}\,(2\,\mathbb{C}_{\rm macro}-\mathbb{C}_{\rm macro})\inv\mathbb{C}_{\rm macro}=2\,\mathbb{C}_{\rm macro}\,,
\end{equation}
i.e.\ simplifying the starting expressions\footnote{Even though the name of the elastic tensors (micro-, meso-, macro-) suggest that the values of $\mathbb{C}_{\rm e}$ should be bounded by $\mathbb{C}_{\rm micro}$ from above and $\mathbb{C}_{\rm macro}$ from below, this is not the case in general. For a very simple unit cell whose structure differs little from the homogeneous cell $\mathbb{C}_{\rm micro}$ is very similar to $\mathbb{C}_{\rm macro}$ implying that $\mathbb{C}_{\rm e}=\mathbb{C}_{\rm micro}\,(\mathbb{C}_{\rm micro}-\mathbb{C}_{\rm macro})\inv\mathbb{C}_{\rm macro}$ tends to infinity. Thus, $\mathbb{C}_{\rm e}$ is not an elastic tensor in the classical sense and very large values impose $\sym P\approx\sym\grad u$.} by $\mathbb{C}_{\rm e}=\mathbb{C}_{\rm micro}$.

We list the numerical values of all parameters obtained via the fitting procedure presented in this section of the relaxed micromorphic model in 
Table~\ref{tab:numericalValuesWithoutCurl}.

\renewcommand{\arraystretch}{1.5}
\begin{table}[ht!]
    \centering
    \begin{tabular}{c|c|c|c||c|c|c}
        $\kappa_{\rm m}$ & $\mu_{\rm m}$ & $\mu_{\rm m}^*$ & $\mu_{\rm c}$ & $\kappa_{\rm e}$ & $\mu_{\rm e}$ & $\mu_{\rm e}^*$\\ \hline
        $[\si{\kPa}]$ & $[\si{\kPa}]$ & $[\si{\kPa}]$ & $[\si{\kPa}]$ & $[\si{\kPa}]$ & $[\si{\kPa}]$ & $[\si{\kPa}]$\\ \hline
        136.5 & 71.73 & 43.76 & 3.81 & 248.2 & 166.1 & 57.48
    \end{tabular}
    \\[2ex]
    \begin{tabular}{c|c|c|c||c|c|c|c}
        $\kappabar_\gamma$ & $\gammabar_1$ & $\gammabar_1^*$ & $\gammabar_2$ & $\kappa_\gamma$ & $\gamma_1$ & $\gamma_1^*$ & $\gamma_2$\\ \hline
        1.732 & 0.124 & 0.523 & 0.053 & 0.635 & 0.407 & 0.086 & 0.173
    \end{tabular}
    \caption{Obtained numerical values of the relaxed micromorphic model without curvature fitted for the metamaterial whose unit cell is given in Figure~\ref{fig:ChartesGeometry}.}
    \label{tab:numericalValuesWithoutCurl}
\end{table}
\renewcommand{\arraystretch}{1}


\subsection{Discussion}

\begin{figure}[ht!]
	\centering
	\includegraphics[width=0.49\textwidth]{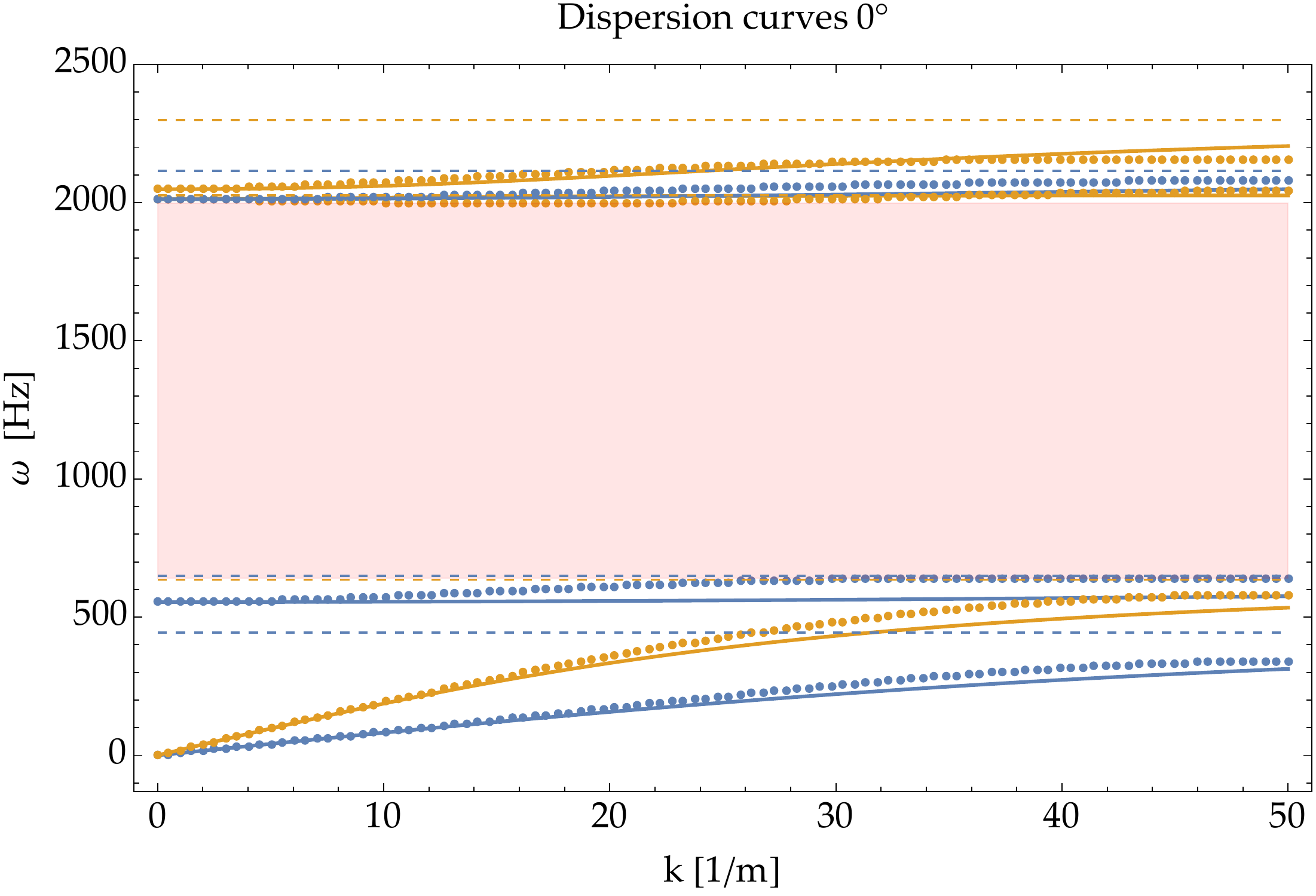}
	\hfill
	\includegraphics[width=0.49\textwidth]{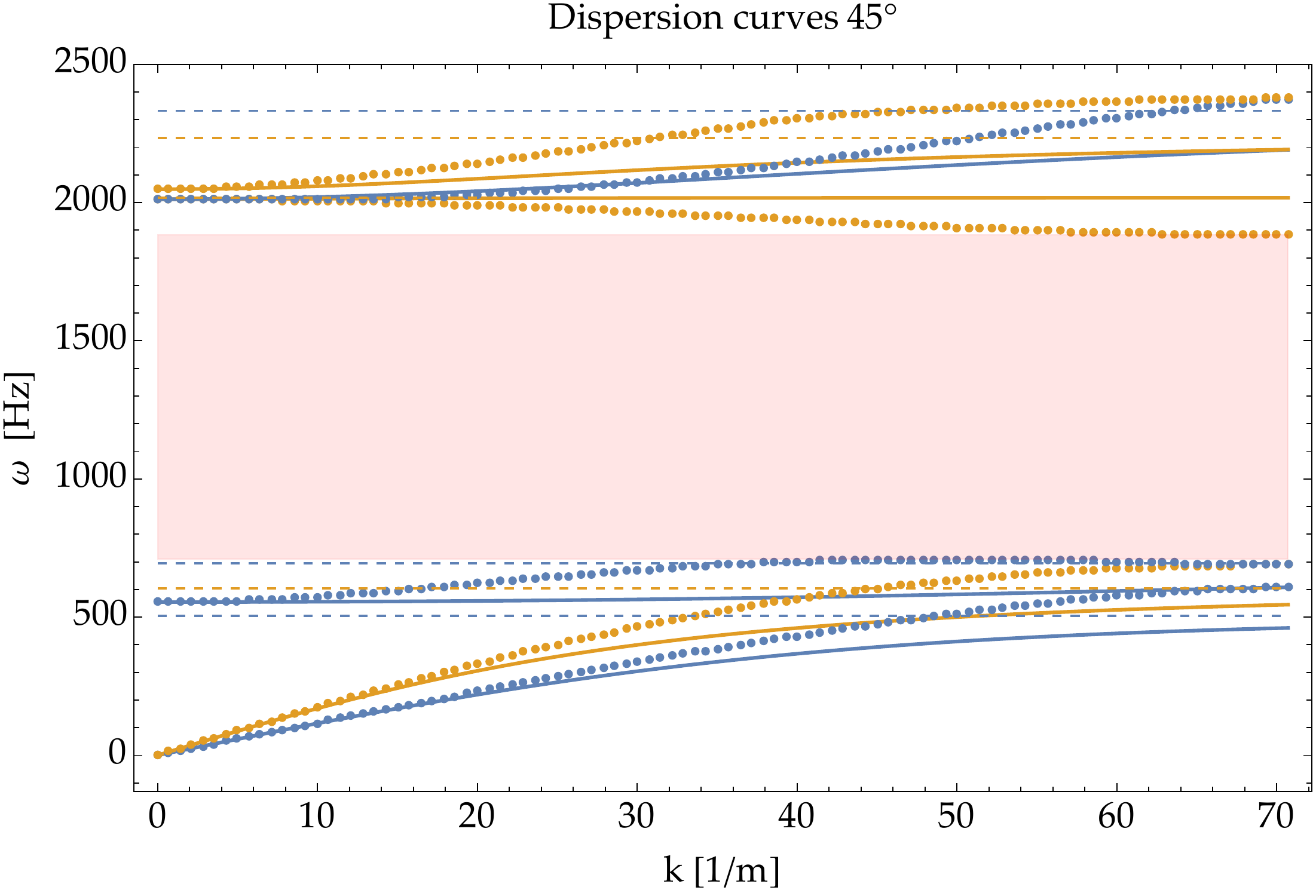}
	\caption{Dispersion curves $\omega(k)$ for 0 degrees (left) and 45 degrees (right) with pressure curves colored in yellow and shear in blue. The dots are the points computed with \comsol\ while the smooth curves show the analytical expression of the dispersion curves for the relaxed micromorphic model without curvature, i.e.\ for $\alpha_1=\beta_1=0$. The value of the curve's horizontal asymptotes are also shown with dashed lines.}\label{fig:fitting1}
\end{figure}

The fitting shown in Figure~\ref{fig:fitting1} behaves well for all frequencies $\omega$ and wavenumber $k$ for zero degrees of incidence but looses some precision for an incidence angle of $45^\circ$ especially for higher values of $k$.
This calls for a further generalization of the relaxed micromorphic model which will be object of following papers. In any case, the achieved overall precision already allows us to explore the dispersive metamaterial's characteristics at a satisfactory level.


The absence of higher-order terms ($\Curl P$ and $\Curl\dot P$) caused the reduction to a single expression $k(\omega)$ describing all three dispersion curves in one, cf.\ equation \eqref{eq:dispRelation}.
Thus for every frequency $\omega$, there is exactly one wavenumber $k$ which may be imaginary if the term inside the root is negative.
For the plots here, we only show $k(\omega)$ where the expression is real-valued and ignore imaginary $k(\omega)$ which arise in the band gap and for higher frequencies.
Moreover, we cannot have two distinct wavenumbers with the same frequency which implies that all curves are monotonic.
For every group of dispersion curves, e.g.\ the three pressure waves for $45^\circ$ incidence, each individual curve is bounded by the others.
Starting with the acoustic curves, their asymptote must be below the cut-off of the lower optic curve of the same type (pressure or shear) while the asymptote of the lower optic curves is bounded from above by the cut-off of the highest optic curve.
In particular, self-intersection between two pressure or two shear curves is not possible with this simplified version of the relaxed micromorphic model.

On the other hand, we observe that for the numerical values from \comsol\ the asymptote of the acoustic shear wave at $45^\circ$ should be slightly higher than the cut-off of the lower optic curve with $581.95~\si{\Hz}$ and $554.61~\si{\Hz}$, respectively.
In addition, assuming that all micro parameters $\kappa_{\rm m},\mu_{\rm m},\mu_{\rm m}^*$ (see Table~\ref{tab:numericalValuesWithoutCurl}) are larger than their corresponding macro counterparts $\kappa_{\rm M},\mu_{\rm M},\mu_{\rm M}^*$, we did not manage to generate decreasing dispersion curves as observed for the lower optic pressure wave for an angle of incidence of $45^\circ$.
This effect can instead be achieved when considering mixed space-time derivatives on the micro-distortion tensor $P$, cf.\ Section~\ref{sec:fitting3}.

When the relative positions of the curves allow to fit the cut-offs and the asymptotes of each curve separately (e.g.\ for an incidence angle of zero degrees shown here) the simplified version of the relaxed micromorphic model used does indeed shows very good results.
We want to emphasize that we only used the limit cases $k=0$ (cut-offs) and $k\to\infty$ (asymptotes) for the fitting procedure but have an appreciable approximation for all values of $k,\omega$.
%
%
%
%
\section{Fitting of the relaxed micromorphic parameters with curvature (with $\Curl P$)}\label{sec:fitting2}

To gain more freedom on the dispersion curves shape, we will now consider the addition of $\Curl P$ resulting in a higher-order polynomial in $k$ which enables up to two distinct values of $k$ for every $\omega\in\Rp$.
At first, we still neglect $\Curl\dot P$ resulting in the slightly reduced characteristic polynomial
\begin{align}
    c_0\.k^2+c_1\.k^4-(c_2+c_3\.k^2+c_4^*\.k^4)\,\omega^2+(c_5+c_6^*\.k^2+c_7^*\.k^4)\,\omega^4-(c_8+c_9^*\,k^2)\,\omega^6&=0\,.\label{eq:dispersionRelationCurlP}
\end{align}
The characteristic polynomial is of second order regarding $k^2$ which, assuming $k>0$, results in two distinct roots
\begin{align}
    k_{1,2}=\frac{-c_0+c_3\.\omega^2-c_6\.\omega^4-c_9^*\.\omega^6\pm\sqrt{-4\.(c_1+c_4^*\.\omega^2+c_7\.\omega^4)(-c_2+c_5\.\omega^4+c_8\.\omega^6)+(c_0-c_3\.\omega^2+c_6^*\.\omega^4+c_9\.\omega^6)^2}}{2\.(c_1+c_4^*\.\omega^2+c_7^*\.\omega^4)}
\end{align}
describing the dispersion relation. The coefficients $c_0,c_1,c_2,c_3,c_4^*,c_5,c_6^*,c_7^*,c_8,c_9^*$ used\footnote{The notation $c_i^*$ still indicates that the coefficient is missing components belonging to the neglected tensor for $\Curl\dot P$ while $c_i$ without an asterisk states that the coefficient is the same of the enhanced relaxed micromorphic model with $\Curl\dot P$.} for the expression above change from $0^\circ$ to $45^\circ$ as well as for considering pressure and shear waves and are included in the Appendix~\ref{app:coefficients2}.

\subsection{Asymptotes}
\label{sec:asymptotes2}

Because the cut-offs are independent of the coefficients with higher order of $k$, they do not change with the addition of $\Curl P$.
Instead, the expressions of the asymptotes hugely differ compared to the expression without $\Curl P$ discussed before.
We only include the terms with the highest order of $k$ available and compute
\begin{align}
    c_1-c_4^*\.\omega^2+c_7^*\.\omega^4=0\qquad&\iff\qquad \omega^4-\frac{c_4^*}{c_7^*}\.\omega^2+\frac{c_1}{c_7^*}=0\notag\\
    &\iff\qquad\omega^2_{1,2}=\frac{c_4^*\pm\sqrt{(c_4^*)^2-4\.c_1c_7^*}}{2\.c_7^*}\,.\label{eq:asymptotesCurlP}
\end{align}
Surprisingly, the asymptotes with $\Curl P$ are significantly simpler because we must only solve a second-order polynomial instead of a third-order polynomial needed for the cut-offs and the asymptotes without $\Curl P$.
We now only have four distinct horizontal asymptotes (two shear and two pressure) in contrast to six before, which means that we must allow that the two curves (one shear and one pressure) will tend to infinity for high values of $k$, and our choice falls on the two highest optic curves.
The same reasoning about the use of the asymptote in Section 
\ref{sec:asymptotes1} is applied here besides for the two highest optic curves that do not have a horizontal asymptote.


\subsection{Fitting}

The fitting procedure starts exactly as described in Section~\ref{sec:fitting}. As a first step, we fit the macro parameters $\kappa_{\rm M},\mu_{\rm M},\mu_{\rm M}^*$ using the slopes of the dispersion curves for $k=0$ which results in the same values as before. Following this, we express the micro-inertia $\kappa_\gamma,\gamma_1,\gamma_1^*,\gamma_2$ as a function of the numerical values of the cut-offs and the remaining independent material parameters.
Overall, we arrive at 4 unknown micro parameters $\kappa_{\rm m},\mu_{\rm m},\mu_{\rm m}^*,\mu_{\rm c}$, 4 unknown inertia parameters $\kappabar_\gamma,\gammabar_1,\gammabar_1^*,\gammabar_2$, and the new elastic parameter $\alpha_1$ belonging to $\Curl P$.

Although the expressions of the asymptotes are different from the ones without the $\Curl P$, we still have the same split between the parameters, resulting in 4 independent parameters for every group of asymptotes, cf.\ Table~\ref{table:dependencyParametersCurl}.

\renewcommand{\arraystretch}{1.5}
\begin{table}[ht!]
\centering
\begin{tabular}{l|ll}
    & pressure &  shear \\ \hline
     $0^\circ$ & $\kappa_{\rm m},\kappabar_\gamma,\mu_{\rm m},\gammabar_1$ & $\mu_{\rm c},\gammabar_2,\mu_{\rm m}^*,\gammabar_1^*$\\
    $45^\circ$ & $\kappa_{\rm m},\kappabar_{\gamma},\mu_{\rm m}^*,\gammabar_1^*$ & $\mu_{\rm c},\gammabar_2,\mu_{\rm m},\gammabar_1$          
\end{tabular}
\caption{Dependence of the asymptotes of the dispersion curves on the free material parameters as function of the direction of propagation ($0^\circ$/$45^\circ$) and type of wave (shear/pressure).}
\label{table:dependencyParametersCurl}
\end{table}
\renewcommand{\arraystretch}{1}

In particular, all 8 expressions are independent on $\alpha_1$ and thus $\Curl P$. Hence, we have no analytic expressions to assign a value to this parameter. Note that the dispersion curves itself (in particular their shape) still depend on $\alpha_1$ but the cut-offs and asymptotes remain independent of this parameter.
Because we miss the higher optic curves asymptotes, we introduce 4 new expressions as a replacement, by considering $k=\frac{1}{L_{\rm c}}$ at $0^\circ$ and $\frac{\sqrt 2}{L_{\rm c}}$ at $45^\circ$ (instead of $k\to\infty$) for the corresponding curves, i.e.\ the highest value for $k$ for which we still have numerical values using Bloch-Floquet analysis.
Note again that $L_{\rm c}$ is always the size of the unit cell.
These \enquote{pseudo-asymptotes} depend on the same parameters as their corresponding acoustic and lower optic curve asymptotes (cf.\ Table~\ref{table:dependencyParametersCurl}) and the additional independent parameter $\alpha_1$.
The analytical expression are again too large to be included here but depend on all the coefficient $c_0,\cdots,c_9^*$ of the dispersion polynomial \eqref{eq:dispersionRelationCurlP}, cf.\ Appendix~\ref{app:coefficients2}.

We list the numerical values of all parameters used for the fitting of the micromorphic model in Table~\ref{tab:numericalValuesWithCurl}.

\renewcommand{\arraystretch}{1.5}
\begin{table}[ht!]
    \centering
    \begin{tabular}{c|c|c|c||c|c|c||c}
        $\kappa_{\rm m}$ & $\mu_{\rm m}$ & $\mu_{\rm m}^*$ & $\mu_{\rm c}$ & $\kappa_{\rm e}$ & $\mu_{\rm e}$ & $\mu_{\rm e}^*$ & $\alpha_1$\\ \hline
        $[\si{\kPa}]$ & $[\si{\kPa}]$ & $[\si{\kPa}]$ & $[\si{\kPa}]$ & $[\si{\kPa}]$ & $[\si{\kPa}]$ & $[\si{\kPa}]$ & $[\si{\kPa}]$\\ \hline
        418.5 & 5010 & 2187 & 3833 & 111.5 & 50.60 & 25.13 & 896.3
    \end{tabular}
    \\[2ex]
    \begin{tabular}{c|c|c|c||c|c|c|c}
        $\kappabar_\gamma$ & $\gammabar_1$ & $\gammabar_1^*$ & $\gammabar_2$ & $\kappa_\gamma$ & $\gamma_1$ & $\gamma_1^*$ & $\gamma_2$\\ \hline
        1.733 & 1.455 & 101.1 & 0.168 & 0.874 & 8.653 & 86.25 & 3.782
    \end{tabular}
    \caption{Numerical values of the relaxed micromorphic model with $\Curl P$ fitted for the metamaterial whose unit cell is given in Figure~\ref{fig:ChartesGeometry}.}
    \label{tab:numericalValuesWithCurl}
\end{table}
\renewcommand{\arraystretch}{1}


\subsection{Discussion}

\begin{figure}[ht!]
	\centering
	\includegraphics[width=0.49\textwidth]{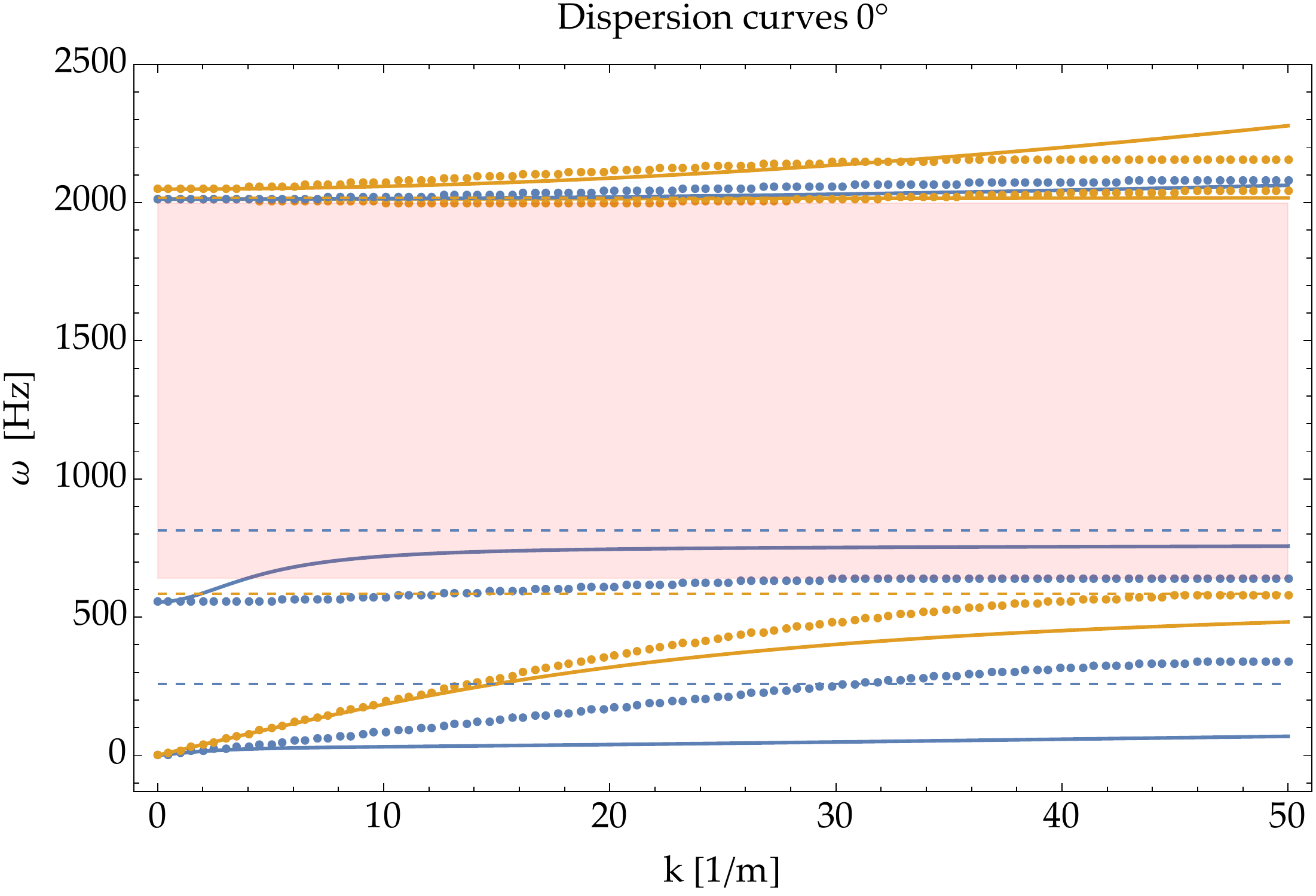}
	\hfill
	\includegraphics[width=0.49\textwidth]{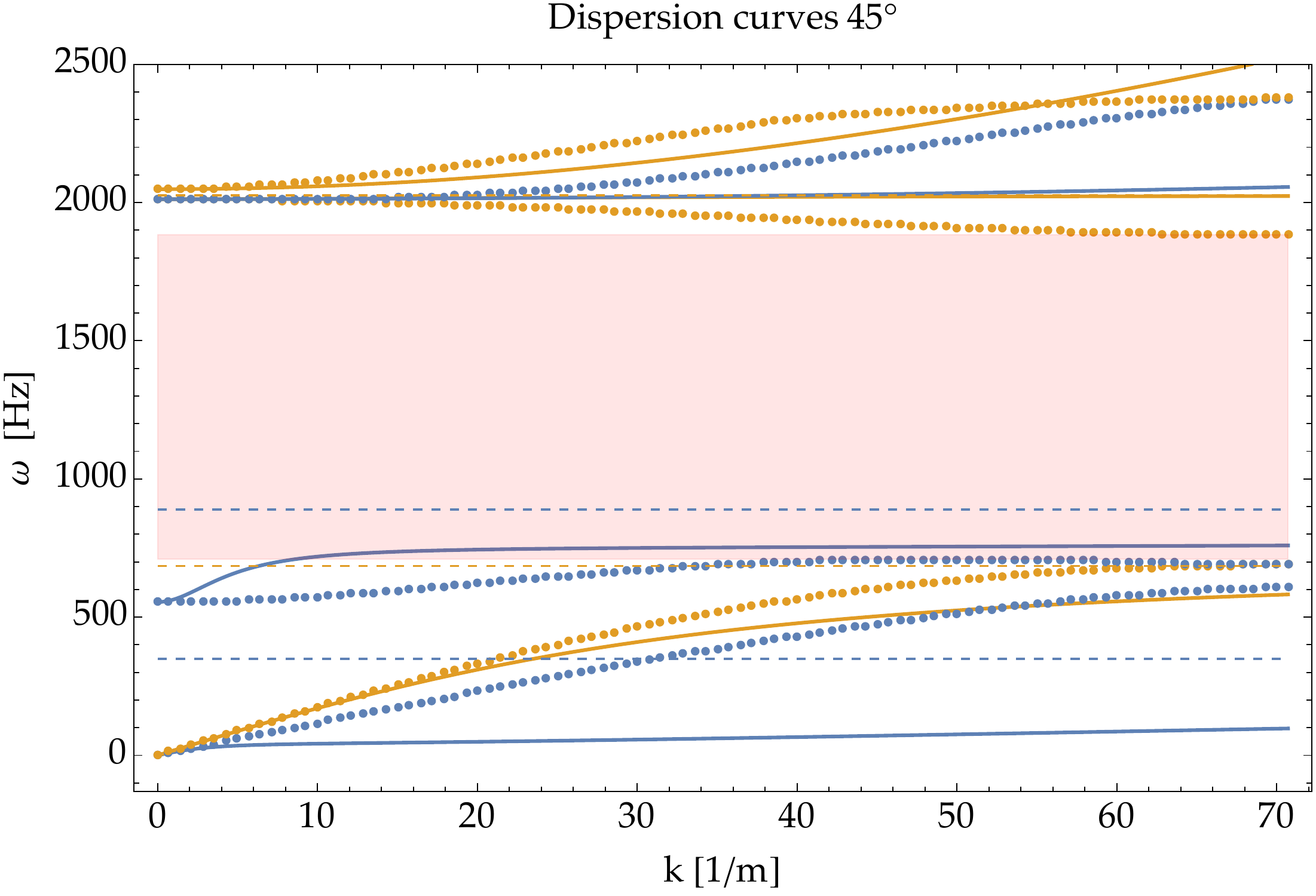}
	\caption{Dispersion curves $\omega(k)$ for 0 degrees (left) and 45 degrees (right) with pressure curves colored in yellow and shear in blue. The dots are the points computed with \comsol\ while the smooth curves show the analytical expression of the dispersion curves for the relaxed micromorphic model for $\beta_1=0$. The value of the curve's horizontal asymptotes are also shown with dashed lines.}\label{fig:fitting2}
\end{figure}

\begin{figure}[ht!]
	\centering
	\includegraphics[width=0.49\textwidth]{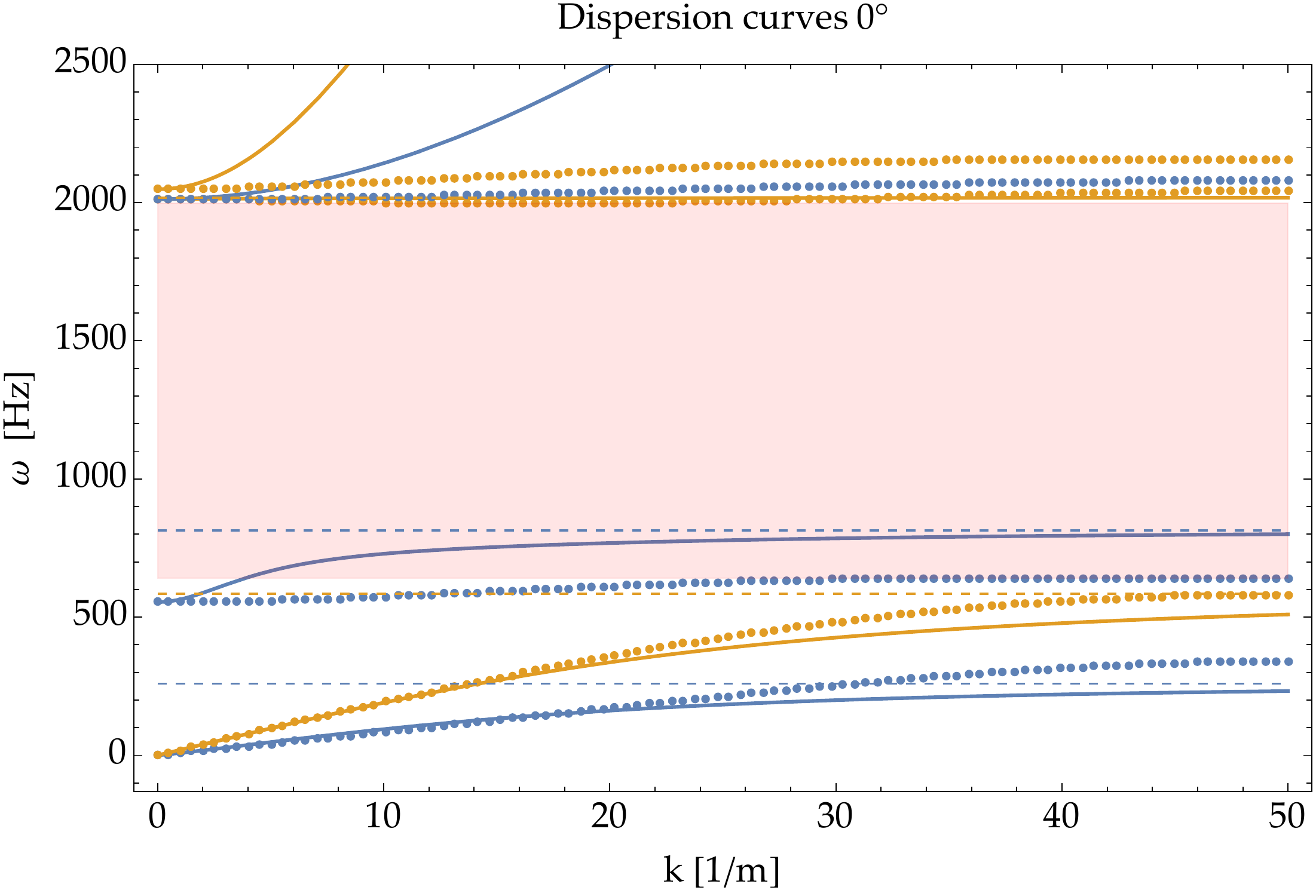}
	\hfill
	\includegraphics[width=0.49\textwidth]{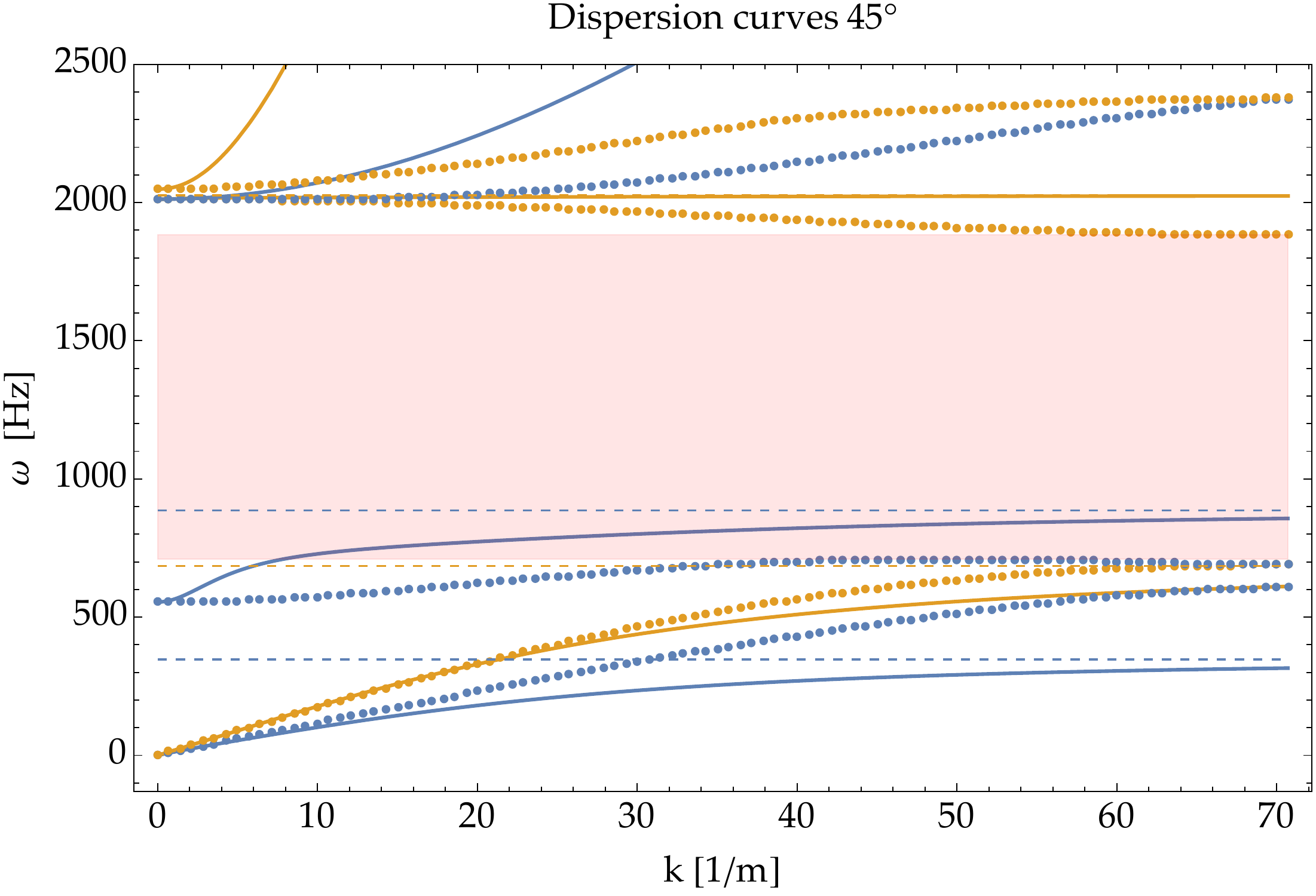}
	\caption{Dispersion curves $\omega(k)$ for 0 degrees (left) and 45 degrees (right) with pressure curves colored in yellow and shear in blue. The dots are the points computed with \comsol\ while the smooth curves show the analytical expression of the dispersion curves for the relaxed micromorphic model for $\beta_1=0$. The value of the curve's horizontal asymptotes are also shown with dashed lines.}\label{fig:fitting2b}
\end{figure}

The fitting shown in Figure~\ref{fig:fitting2} including the curvature $\Curl P$ is worse compared to the one without, cf.\ Figure~\ref{fig:fitting1} and Table~\ref{tab:numericalValuesWithCurl}.
This is mainly because we lost the higher optic curves asymptotes. Additionally, the shape of the curves which comes automatically by fitting all material parameters using only the cut-offs and asymptotes does not match properly with the data computed numerically with \comsol.
Most importantly, the fitting with $\Curl P$ shows the additional challenge of asymptotes that are approached at a very high wavenumber $k\gg\frac{1}{L_{\rm c}}$ resulting in a poor fit for values of $k$ between zero and the size of the unit cell, even if the slopes of the acoustic curves close to zero are well fitted.
In particular, the acoustic shear wave is notably too slow at approaching its limit.
Increasing $\alpha_1$ substantially helps for lower frequencies but causes a much higher "pseudo-asymptote", i.e.\ a worse fit for higher frequencies.
To show this, we did a second fit only using the asymptotes of the acoustic and lower optic curves with $\alpha_1$ as an independent parameter set by hand, cf.\ Figure~\ref{fig:fitting2b}.

Most other values remain at the same magnitude but are slightly higher, cf,\ Table~\ref{tab:numericalValuesWithCurl2}. In future works, we will consider an enhanced relaxed micromorphic model to better describe these effects.

\renewcommand{\arraystretch}{1.5}
\begin{table}[ht!]
    \centering
    \begin{tabular}{c|c|c|c||c|c|c||c}
        $\kappa_{\rm m}$ & $\mu_{\rm m}$ & $\mu_{\rm m}^*$ & $\mu_{\rm c}$ & $\kappa_{\rm e}$ & $\mu_{\rm e}$ & $\mu_{\rm e}^*$ & $\alpha_1$\\ \hline
        $[\si{\kPa}]$ & $[\si{\kPa}]$ & $[\si{\kPa}]$ & $[\si{\kPa}]$ & $[\si{\kPa}]$ & $[\si{\kPa}]$ & $[\si{\kPa}]$ & $[\si{\kPa}]$\\ \hline
        354.7 & 5010 & 2234 & 3902 & 117.1 & 50.60 & 25.12 & \num{6e7}
    \end{tabular}
    \\[2ex]
    \begin{tabular}{c|c|c|c||c|c|c|c}
        $\kappabar_\gamma$ & $\gammabar_1$ & $\gammabar_1^*$ & $\gammabar_2$ & $\kappa_\gamma$ & $\gamma_1$ & $\gamma_1^*$ & $\gamma_2$\\ \hline
        0.904 & 2.388 & 102.0 & 1.072 & 0.779 & 8.653 & 87.79 & 3.863
    \end{tabular}
    \caption{Second set of alternative numerical values of the relaxed micromorphic model with $\Curl P$ fitted for the metamaterial whose unit cell is given in Figure~\ref{fig:ChartesGeometry}.}
    \label{tab:numericalValuesWithCurl2}
\end{table}
\renewcommand{\arraystretch}{1}
%
%
%
%
\section{Fitting of the relaxed micromorphic parameters with enhanced kinetic energy (with $\Curl\dot P$)}\label{sec:fitting3}

We now include $\Curl\dot P$ which reintroduces the third asymptote by considering the full dispersion polynomial

\begin{equation}
    c_0\.k^2+c_1\.k^4-(c_2+c_3\.k^2+c_4\.k^4)\,\omega^2+(c_5+c_6\.k^2+c_7\.k^4)\,\omega^4-(c_8+c_9\,k^2+c_{10}\.k^4)\,\omega^6=0\,,\label{eq:dispersionRelationCurldotP}
\end{equation}
with coefficients $c_0,\cdots,c_{10}$ depending on all the material parameters described before and $\beta_1$ belonging to $\Curl\dot P$, cf.\ Appendix~\ref{app:coefficients3}.


\subsection{Asymptotes}

Again, the cut-offs are independent on the coefficients with higher order of $k$ and thus they do not change with respect to the two previous cases. For the asymptotes we only consider the terms with the highest order of $k$ available and compute
\begin{equation}
    c_1-c_4\.\omega^2+c_7\.\omega^4-c_{10}\.\omega^6=0\qquad\iff\qquad\omega^6-\frac{c_7}{c_{10}}\.\omega^4+\frac{c_4}{c_{10}}\,\omega^2-\frac{c_1}{c_{10}}\,.
    \label{eq:asymptotesCurlPDot}
\end{equation}
We have again three asymptotes (the roots of a third order polynomial) which in general causes the analytical expressions to be impractical rather quickly. However, in this case it is possible to find one root by hand
\begin{equation}
    c_1-c_4\.\omega^2+c_7\.\omega^4-c_{10}\.\omega^6=0\qquad\iff\qquad\left(1-\rho L_{\rm c}^2\.\frac{\beta_1}{\alpha_1}\.\omega^2\right)(c_1-c_4^*\.\omega^2+c_7^*\.\omega^4)
    \label{eq:asymptotesCurlPdot}
\end{equation}
with $c_4^*$ and $c_7^*$ from the dispersion relation \eqref{eq:dispersionRelationCurlP} without $\Curl\dot P$ which holds because
\begin{equation}
    \rho L_{\rm c}^2\.\frac{\beta_1}{\alpha_1}\.c_1=c_4-c_4^*\qquad\text{and}\qquad
    \rho L_{\rm c}^2\.\frac{\beta_1}{\alpha_1}\.c_4^*=c_7-c_7^*\qquad\text{and}\qquad
    \rho L_{\rm c}^2\.\frac{\beta_1}{\alpha_1}\.c_7^*=c_{10}
\end{equation}
for all combinations of shear/pressure and $0^\circ$/$45^\circ$ angle of incidence. This is remarkable because it allows us to use the same analytical expressions for the acoustic and lower optic asymptotes from the calculations without $\Curl\dot P$ of Section~\ref{sec:fitting2} while the general expressions of the dispersion curves differ because of the addition of new terms. Secondly, the third asymptote
\begin{equation}
    \omega_3=\sqrt{\frac{\alpha_1}{\rho L_{\rm c}^2\.\beta_1}}\label{eq:thirdAsymptoteCurlPDot}
\end{equation}
is identical for shear and pressure and invariant under change of angle of incidence.
The same reasoning about the use of the asymptote in Section
\ref{sec:asymptotes1} is applied here.

\subsection{Fitting}

The fitting starts as described in the sections before, using cut-offs and the slope of acoustic waves for $k=0$ to reduce the number of independent parameters. Here we arrive at the same 4 unknown micro parameters $\kappa_{\rm m},\mu_{\rm m},\mu_{\rm m}^*,\mu_{\rm c}$ and 4 unknown inertia parameters $\kappabar_\gamma,\gammabar_1,\gammabar_1^*,\gammabar_2$ with the additional elastic parameter $\alpha_1,\beta_1$ belonging to $\Curl P$ and $\Curl\dot P$, respectively. For the latter, we choose the acoustic pressure and lower optic shear curves as the one with the superimposed asymptote. The former are determined numerically using the remaining asymptotes (8 in total), see also Table~\ref{table:dependencyParametersCurldot}.

\renewcommand{\arraystretch}{1.5}
\begin{table}[ht!]
\centering
\begin{tabular}{l|ll}
    & pressure &  shear \\ \hline
     $0^\circ$ & $\kappa_{\rm m},\kappabar_\gamma,\mu_{\rm m},\gammabar_1$ & $\mu_{\rm c},\gammabar_2,\mu_{\rm m}^*,\gammabar_1^*$\\
    $45^\circ$ & $\kappa_{\rm m},\kappabar_{\gamma},\mu_{\rm m}^*,\gammabar_1^*$ & $\mu_{\rm c},\gammabar_2,\mu_{\rm m},\gammabar_1$\\
    superimposed & $\alpha_1,\beta_1$ & $\alpha_1,\beta_1$
\end{tabular}
\caption{Dependence of the asymptotes of the dispersion curves on the free material parameters as function of the direction of propagation ($0^\circ$/$45^\circ$) and type of wave (shear/pressure).}
\end{table}
\label{table:dependencyParametersCurldot}
\renewcommand{\arraystretch}{1}

However, because the analytical expression \eqref{eq:thirdAsymptoteCurlPDot} of these four highest asymptotes are identical, while in general, the numerical values from \comsol\ can be distinct, we will arrive at the average of these four asymptotes. Moreover, we can only fix the ratio $\frac{\alpha_1}{\beta_1}$ which leaves us with one last free parameter, i.e.\ independent of the values of all cut-offs and asymptotes, which we fit by hand.

We list the numerical values of all parameters used in the micromorphic model in Table~\ref{tab:numericalValuesWithCurldot}.

\renewcommand{\arraystretch}{1.5}
\begin{table}[ht!]
    \centering
    \begin{tabular}{c|c|c|c||c|c|c||c}
        $\kappa_{\rm m}$ & $\mu_{\rm m}$ & $\mu_{\rm m}^*$ & $\mu_{\rm c}$ & $\kappa_{\rm e}$ & $\mu_{\rm e}$ & $\mu_{\rm e}^*$ & $\alpha_1$\\ \hline
        $[\si{\kPa}]$ & $[\si{\kPa}]$ & $[\si{\kPa}]$ & $[\si{\kPa}]$ & $[\si{\kPa}]$ & $[\si{\kPa}]$ & $[\si{\kPa}]$ & $[\si{\Pa}]$\\ \hline
        8806 & 57.09 & 27.67 & 13.16 & 88.95 & 408.8 & 243.4 & 37.34
    \end{tabular}
    \\[2ex]
    \begin{tabular}{c|c|c|c||c|c|c|c||c}
        $\kappabar_\gamma$ & $\gammabar_1$ & $\gammabar_1^*$ & $\gammabar_2$ & $\kappa_\gamma$ & $\gamma_1$ & $\gamma_1^*$ & $\gamma_2$ & $\beta_1$\\ \hline
        0.292 & 0.501 & 0.569 & 0.205 & 14.67 & 0.797 & 0.296 & 0.464 & \num{8.405e-4}
    \end{tabular}
    \caption{Numerical values for the relaxed micromorphic model with enhanced kinetic and potential energy terms ($\Curl P$, $\Curl\dot P$) for the metamaterial whose unit cell is given in Figure~\ref{fig:ChartesGeometry}.}
    \label{tab:numericalValuesWithCurldot}
\end{table}
\renewcommand{\arraystretch}{1}


\subsection{Discussion}

\begin{figure}[ht!]
	\centering
	\includegraphics[width=0.49\textwidth]{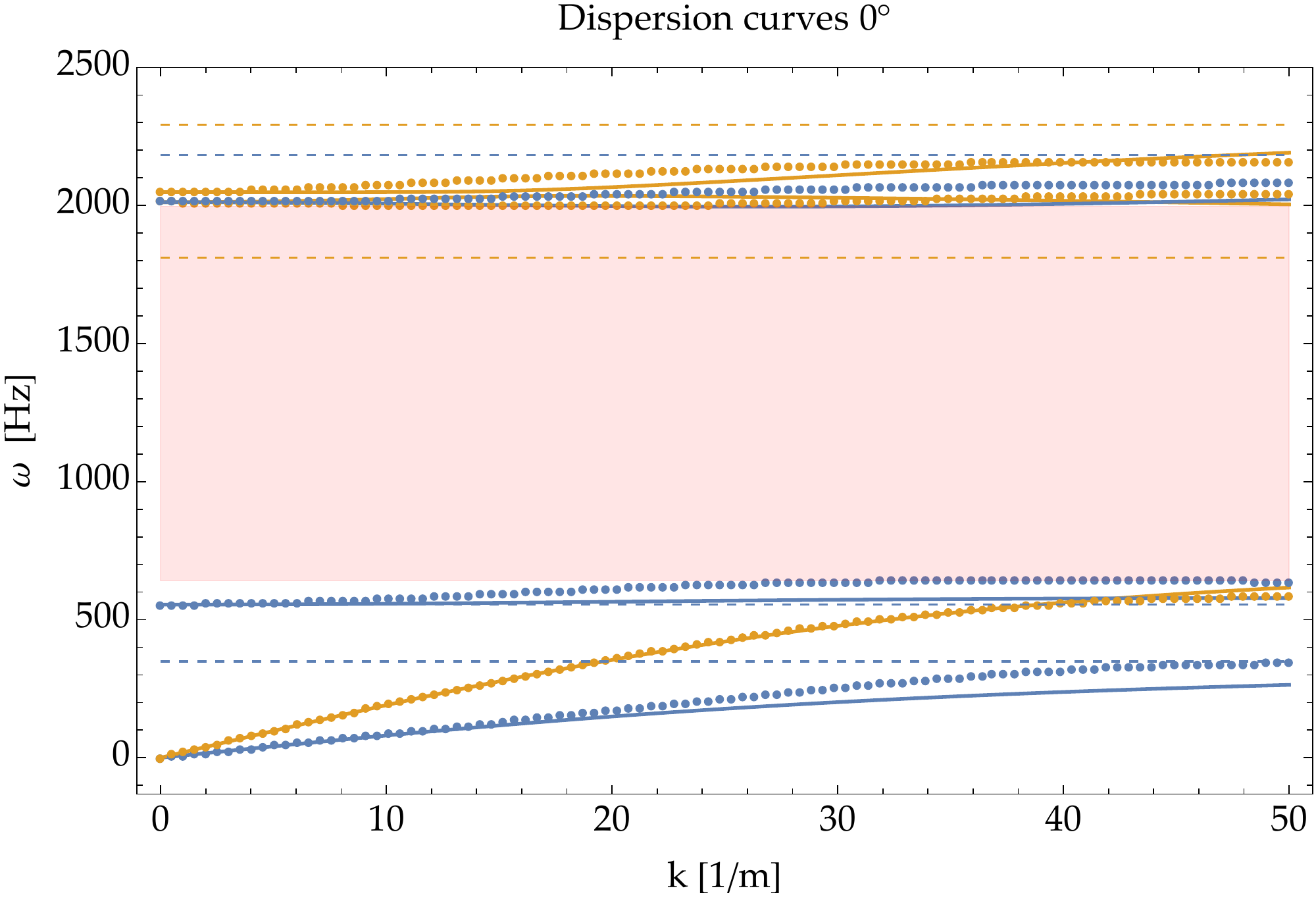}
	\hfill
	\includegraphics[width=0.49\textwidth]{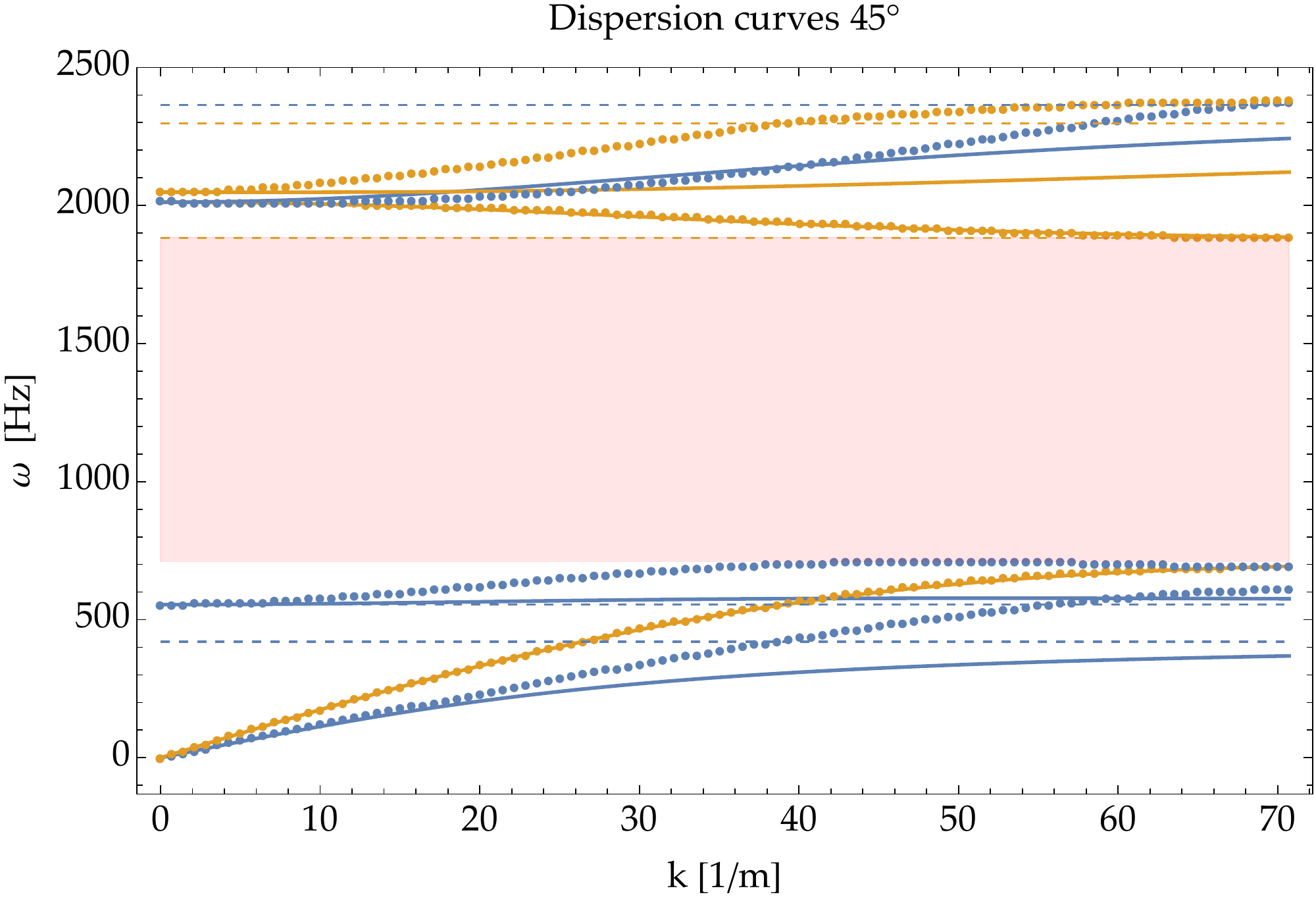}
	\caption{Dispersion curves $\omega(k)$ for 0 degrees (left) and 45 degrees (right) with pressure curves colored in yellow and shear in blue. The dots are the points computed with \comsol\ while the smooth curves show the analytical expression of the dispersion curves for the relaxed micromorphic model. The value of the curve's horizontal asymptotes are also shown with dashed lines.}\label{fig:fitting3}
\end{figure}

The introduction of $\Curl\dot P$ increases the quality of fitting again (see Figure \ref{fig:fitting3}) with the recovery of the higher optic curve asymptote while the remaining free parameter $\alpha_1,\beta_1$ can improve the shape of the curves.
In particular, we can fit a dispersion curve with negative group velocity perfectly, the lower optic pressure curve for 45 degree, for the first time.

The fitting of the other curves is slightly worse compared to the calculations without curvature terms, and it is mainly due to the degenerated shape of $\Curl\dot P$ and $\Curl P$ in a plane problem, resulting in an isotropic behaviour with only a single independent parameter.
This causes four asymptote to coincide.

We won't discuss in a fourth section the fitting with an enhanced kinetic energy with $\Curl\dot P$ but without the curvature $\Curl P$, since it leads to a degenerate dispersion polynomial $p(k,\omega)$ similar to the one described in Section~\ref{sec:fitting2} resulting in only four configurable horizontal asymptotes for shear and pressure each.
However, while we lost the higher optic asymptotes by adding $\Curl P$ without $\Curl\dot P$, the case with $\Curl\dot P$ and without curvature $\Curl P$ instead forces the limits of the one curve to zero as it can be seen as a limit case of expression \eqref{eq:thirdAsymptoteCurlPDot}, i.e.\ $\alpha_1\to 0$.
%
%
%
%
\section{Summary of the obtained results}
Comparing the results of Section~\ref{sec:fitting1}, Section~\ref{sec:fitting2}, and Section~\ref{sec:fitting3}, the first approach which does not include $\Curl P$ and $\Curl\dot P$ shows the best agreement to the numerical values from \comsol\ overall.
At the same time, it must be noted that this simplified model without any curvature terms is only capable of describing monotonic dispersion curves with three disjointed domains for the three pressure waves and three disjointed domains for the three shear waves (crossing between pressure and shear waves is allowed while it is not allowed between curves of the same type).
The latter property can be easily deduced from equation $\eqref{eq:dispRelation}_{2}$ which guarantees that for each value of $\omega$ there can only be one value of $k$.

In order to remove this constraint, the dispersion relation $p(k,\omega)=0$ must contain higher order terms of the wavenumber $k$ to allow the overlapping of the domains of curves of the same kind.
To this aim, we started considering the full relaxed micromorphic model (with $\Curl P$) and then augmented it with a new inertia term ($\Curl\dot P$).
The comparison of the analytic expression of the asymptotes with or without $\Curl\dot P$ shown in equation \eqref{eq:asymptotesCurlPdot} suggests that, when a new term is added to the elastic energy density, it is always better to include the corresponding dynamic part as well.
In particular, considering $\Curl P$ without its counterpart $\Curl\dot P$ or vice verse causes missing terms in the dispersion polynomial $p(k,\omega)=0$ resulting in fewer horizontal asymptotes.

While the model without curvature terms gives the best quantitative agreement (see Figure \ref{fig:fitting1}), the augmented relaxed micromorphic model still gives a good agreement and opens up the important possibility of decreasing modes with negative group velocity, cf.\ Figure \ref{fig:fitting3}.
Note that $\Curl P$ and $\Curl\dot P$ degenerate for planar problems: for both terms, it just remains a single independent constitutive parameter ($\alpha_1$ and $\beta_1$) restricting the class of symmetry for their constitutive tensors to the isotropic one.

We want to emphasize that the main focus of this work is not the result of the fitting of the three different approaches per se, but the semi-analytical fitting algorithm itself and the underlying consistency of the relaxed micromorphic model with respect to the material properties and some of the geometrical characteristic of the metamaterial that it represents.
Using the complex but analytically defined expressions of the asymptotes, we can find a numerical fit of all material parameters by only giving the numerical values computed with \comsol\ and the apparent mass density $\rho$ of the unit cell.
Note that we only use the cut-offs $k=0$ and asymptotes $k\to\infty$ for calculating the material parameters while the shape of the curves for intermediate values of $k$ comes automatically.

The routine is completely written with \mathematica\ allowing us to use symbolic calculations.
The essential part of the fitting procedure uses the inbuilt algorithm \textit{NMinimize} (with the Method \textit{RandomSearch}) to minimize the mean square error of the asymptotes between the relaxed micromorphic model and the numerical values of the finite element approach in \comsol.
Therefore, in general, if a local minimum is found, it is not guaranteed that it corresponds to a global optimum as well.

\begin{figure}[ht!]
	\centering
	\includegraphics[width=0.49\textwidth]{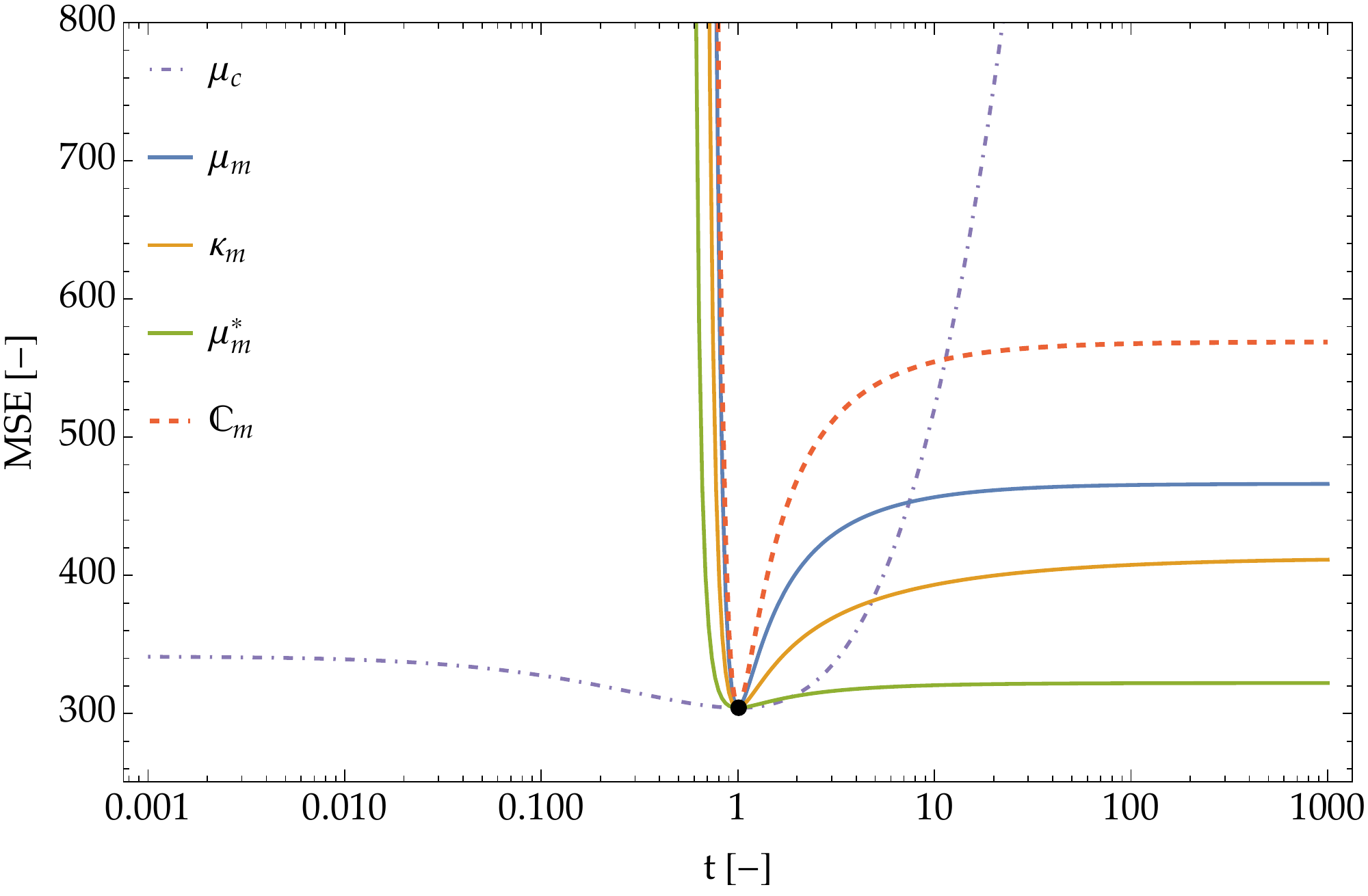}
	\hfill
	\includegraphics[width=0.49\textwidth]{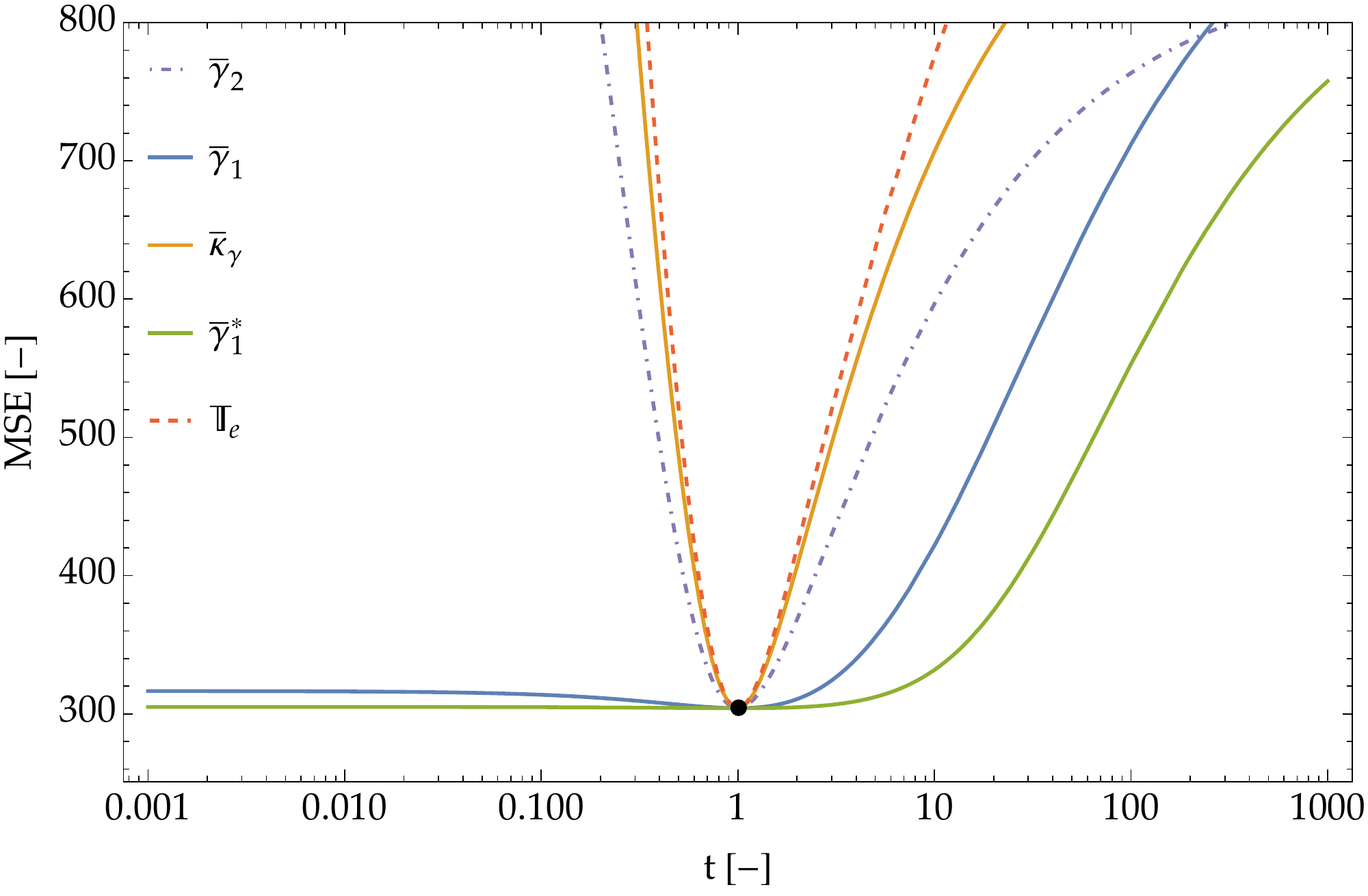}
	\caption{Mean squared error between the analytic expressions from the relaxed micromorphic model and their corresponding numerical values from \comsol while changing the scaling coefficient $t$ for different material parameters. Left: the static parameters $\kappa_{\rm m}, \mu_{\rm m}, \mu_{\rm m}^*, \mu_{\rm c}$ are colored in orange, blue, green and purple while the dashed red line shows the impact of scaling $\mathbb{C}_{\rm micro}$ as a whole. Right: the dynamic parameters $\kappabar_\gamma, \gammabar_1, \gammabar_1^*, \gammabar_2$ are colored in orange, blue, green and purple while the dashed red line shows the impact of scaling $\mathbb{T}_{\rm e}$ as a whole.}
	\label{fig:changeValues}
\end{figure}

In order to better understand the nature of the minimization problem, we visualize the impact of each parameter thanks to Figure~\ref{fig:changeValues} for the case without $\Curl P$ and $\Curl\dot P$ of Section~\ref{sec:fitting}: given a set of material constants values, we move one parameter at the time and plot the impact on the error for the asymptotes.

It is crucial to state here, that the expression of the coefficients shown in Appendix~\ref{app:coefficients}-\ref{app:coefficients3} are a priori constrained by the relations described in Section~\ref{sec:fitting} which reduces the number of independent parameters.
This guarantees that the values of the cut-offs will not change if the micro parameters change because the corresponding micro-inertia parameters $\kappa_\gamma, \gamma_1, \gamma_1^*, \gamma_2$ are automatically scaled as well.

When considering the impact of the different independent parameters on the fitting error of the asymptotes, visible dissimilarity arises.
While most coefficients show a simple behaviour with one distinct minimum, the impact on the dispersion curves of the parameters $\mu_{\rm m}^*$ and $\gammabar_1^*$ is very small.
In addition to the numerical difficulty caused by a vanishing gradient, it suggests finding additional quantities to constrain these material parameters, e.g.\ use a static test to fix $\mu_{\rm m}^*$ beforehand.
Overall, the minimum problem behaves remarkably well given its simple shape, which is in accordance with the fact that the fitting procedure always converges to very similar values of the parameters regardless of their initial guess.
%
%
%
%
\section{Conclusion and perspectives}
We presented an inertia-augmented relaxed micromorphic model that enhances the model proposed in \cite{aivaliotis_microstructure-related_2019,aivaliotis_frequency_2020,rizzi_exploring_2021,rizzi2021boundary,rizzi2021metamaterial} via the addition of a term $\Curl \dot{P}$ in the kinetic energy.
We then used this model to describe the dynamical behavior of a labyrinthine metamaterial and compared its predictability when setting all curvature terms to zero, when setting only $\Curl \dot{P}$ to zero and when considering the full model (including $\Curl P$ and $\Curl \dot{P}$).
The model’s efficiency is tested by comparing the obtained dispersion curves to the ones issued via a standard Bloch-Floquet analysis.
We find that the model without curvature terms gives the best average behavior (when considering the whole set of frequencies and wave numbers).
However, this reduced model shows two main drawbacks: i) it cannot give rise to curves with negative group velocity and ii) the horizontal asymptote of each curve is bounded by the cutoff of the following one.
This implies a limitation of the fitting quality at higher wave numbers for some directions of propagation.
To remove these constraints, the full model (including $\Curl P$ and $\Curl \dot{P}$) can be used.
The fitting that we obtained with the full model remains of good quality (although precision is sometimes lowered point-wise) and we could perfectly describe (for the first time in a micromorphic framework) a higher-frequency mode with negative group velocity.
The constraint concerning the asymptotes’ boundedness is also removed in the full model, but a perfect fitting at higher wave-numbers cannot still be achieved due to a lack of extra microscopic degrees of freedom that are needed (at least for some directions) for a perfect fitting at wavelengths approaching the size of the unit cell.

We also tested the model’s performances by considering only $\Curl P$ and not $\Curl \dot{P}$.
The fitting quality is visibly worsening, thus suggesting that, when introducing a term in the strain energy, its dynamical counterpart in the kinetic energy should be always considered as well.

In addition to the fitting comparison presented in the present paper, one main result that we present here is the fitting procedure itself that has been automatized to a big extent by asking only the cut-offs and asymptotes to be imposed a priori.
This has been done by imposing the exact value of the cut-offs and minimizing the asymptotes' mean square error compared to the exact numerical values issued via Boch-Floquet analysis.
The rest of the curves’ fitting follows directly.

The routine for the fitting procedure is written with \mathematica allowing us to use symbolic calculations.
The essential part of the routine uses the built-in algorithm \textit{NMinimize} (with the Method \textit{RandomSearch}) to minimize the asymptote’s mean square error.
We finally checked whether the found local minimum is also a global minimum.
We found that all the elastic parameters achieve a global minimum in the computed configuration, but $\mu_{\rm m}^*$ and $\gammabar_1^*$ seem to have little effect on the overall fitting after a certain threshold.
This result seems to suggest that the value of these parameters should be eventually fixed beforehand with the help of extra independent static and/or dynamic tests.

Based on the findings of this paper, we will briefly present some insight that will give directions to the follow-up research:
\begin{itemize}
    \item Further enhance the relaxed micromorphic model via the addition of extra microscopic degrees of freedom to increase its precision at very small wavelengths (approaching the unit cell’s size);
    \item Design complex large-scale meta-structures that control elastic energy using the new labyrinthine metamaterial as a basic building block. This design would not be otherwise possible due to the huge number of degrees of freedom resulting from the meshing of all the tiny elements contained in the labyrinthine unit cells;
    \item Study negative refraction phenomena in meta-structures including the new labyrinthine metamaterial as a basic building block;
    \item Design complex structures for wave control simultaneously including the different metamaterials that were characterized via the relaxed micromorphic model until now.
\end{itemize}
%
%
%
%
\small
\subsubsection*{Acknowledgements}
Angela Madeo, Gianluca Rizzi and Jendrik Voss acknowledge support from the European Commission through the funding of the ERC Consolidator Grant META-LEGO, N$^\circ$ 101001759.
Angela Madeo and Gianluca Rizzi acknowledge funding from the French Research Agency ANR, “METASMART” (ANR-17CE08-0006).
Patrizio Neff acknowledges support in the framework of the DFG-Priority Programme 2256 ``Variational Methods for Predicting Complex Phenomena in Engineering Structures and Materials'', Neff 902/10-1, Project-No. 440935806.

\footnotesize
\printbibliography

\footnotesize
\begin{appendix}
\end{appendix}
\section{Most general 4th order tensor belonging to the tetragonal symmetry class }\label{app:SymSkewCoup}
Considering the following quadratic form
\begin{equation}
    Y=\iprod{\mathbb{L}\.D,D}\,,
    \label{eq:quadForm}
\end{equation}
where $\mathbb{L}$ is a 4th order tensor and $D$ is a 2nd order one, the most general form of $\mathbb{L}$ if it belongs to the tetragonal symmetry class written in Voigt notation is 
\begin{equation}
    L=\left(
\begin{array}{ccccccccc}
 \ell_{1} & \ell_{5} & \ell_{6} & 0 & 0 & 0 & 0 & 0 & 0 \\
 \ell_{5} & \ell_{1} & \ell_{6} & 0 & 0 & 0 & 0 & 0 & 0 \\
 \ell_{6} & \ell_{6} & \ell_{2} & 0 & 0 & 0 & 0 & 0 & 0 \\
 0 & 0 & 0 & \ell_{3}+\ell_{7}+\ell_{9} & \ell_{3}-\ell_{7} & 0 & 0 & 0 & 0 \\
 0 & 0 & 0 & \ell_{3}-\ell_{7} & \ell_{3}+\ell_{7}-\ell_{9} & 0 & 0 & 0 & 0 \\
 0 & 0 & 0 & 0 & 0 & \ell_{3}+\ell_{7}+\ell_{9} & \ell_{3}-\ell_{7} & 0 & 0 \\
 0 & 0 & 0 & 0 & 0 & \ell_{3}-\ell_{7} & \ell_{3}+\ell_{7}-\ell_{9} & 0 & 0 \\
 0 & 0 & 0 & 0 & 0 & 0 & 0 & \ell_{4}+\ell_{8} & \ell_{4}-\ell_{8} \\
 0 & 0 & 0 & 0 & 0 & 0 & 0 & \ell_{4}-\ell_{8} & \ell_{4}+\ell_{8}\\
\end{array}
\right)\,,
\end{equation}
where the order of the element of the vector associated with the quadratic form~\ref{eq:quadForm} is
\begin{equation}
    d=
    \left(
    D_{11},D_{22},D_{33},D_{23},D_{32},D_{13},D_{31},D_{12},D_{21}
    \right)^{\rm T} \,.
\end{equation}
If we now split the tensor $D$ in its symmetric and skew-symmetric part, the corresponding vector in Voigt notation are
\begin{equation}
    d_{s}=
    \left(
    D_{11},D_{22},D_{33},D_{23}+D_{32},D_{13}+D_{31},D_{12}+D_{21}
    \right)^{\rm T} \,,
    \qquad
    d_{a}=
    \frac{1}{2}
    \left(
    D_{23}-D_{32},D_{13}-D_{31},D_{12}-D_{21}
    \right)^{\rm T} \,.
\end{equation}
Because of the class of symmetry considered, it is necessary to take into account a mixed constitutive matrix that couples the symmetric and skew-symmetric part of $D$ in order to build back the quadratic form $Y$
\begin{equation}
    Y=\iprod{ \mathbb{L} \. D,D}
    =\iprod{ L \. d,d}
    =\iprod{ L_{\rm s} \. d_{\rm s},d_{\rm s}}
    + \iprod{ L_{\rm a} \. d_{\rm a},d_{\rm a}}
    + 2\iprod{ L_{\rm mix} \. d_{\rm s},d_{\rm a}}
\end{equation}
where
\begin{align}
    L_{\rm s}=
    \left(
    \begin{array}{cccccc}
     \ell_{1} & \ell_{5} & \ell_{6} & 0 & 0 & 0 \\
     \ell_{5} & \ell_{1} & \ell_{6} & 0 & 0 & 0 \\
     \ell_{6} & \ell_{6} & \ell_{2} & 0 & 0 & 0 \\
     0 & 0 & 0 & \ell_{3} & 0 & 0 \\
     0 & 0 & 0 & 0 & \ell_{3} & 0 \\
     0 & 0 & 0 & 0 & 0 & \ell_{4} \\
    \end{array}
    \right)
    \,,
    \quad
    L_{\rm a}=
    \left(
    \begin{array}{ccc}
     4 \ell_{7} & 0 & 0 \\
     0 & 4 \ell_{7} & 0 \\
     0 & 0 & 4 \ell_{8} \\
    \end{array}
    \right)
    \,,
    \quad
    L_{\rm mix}=
    \left(
    \begin{array}{cccccc}
     0 & 0 & 0 & \ell_{9} & 0 & 0 \\
     0 & 0 & 0 & 0 & \ell_{9} & 0 \\
     0 & 0 & 0 & 0 & 0 & 0 \\
    \end{array}
    \right)\, .
\end{align}
Only the coefficient $\ell_{9}$ couples the symmetric and skew-symmetric part of $D$, and it produce works just for out of plane deformation if a plane strain problem ($x_1 O x_2$) in considered, which means that it is non involved in the constitutive relations for both the gradient of the displacement $\nabla u$ or the micro-distortion tensor $P$, while it is for the $\Curl P$ or for $\Curl \dot P$.
Nevertheless, the quadratic form $Y$ when $D=\Curl P$ (see equation \eqref{eq:curl_shape}) under a plane strain hypothesis is
\begin{equation}
    Y=(\ell_{3}+\ell_{7}+\ell_{9})\left((\Curl P)_{13}^2+(\Curl P)_{23}^2\right)
\end{equation}
which depends on just one cumulative coefficient.
This makes the coupling coefficient $\ell_9$ and the skew-symmetric coefficient $\ell_7$ redundant.

\section{Coefficients for the dispersion curves without $\Curl P$}\label{app:coefficients}

For the relaxed micromorphic model without curvature, we arrived \eqref{eq:dispRelation} at the dispersion polynomial
\begin{equation}
	c_0\.k^2-(c_2+c_3^*\.k^2)\,\omega^2+(c_5+c_6^*\.k^2)\,\omega^4-(c_8+c_9^*\,k^2)\,\omega^6
\end{equation}
We list the coefficients $(c_0,c_2,c_3^*,c_5,c_6^*,c_8,c_9^*)^T$.
\begin{equation}
	\matr{
		4 \left(\mu _{\rm e} \kappa _{\rm m} \mu _{\rm m}+\kappa _{\rm e} \left(\kappa _{\rm m} \left(\mu _{\rm e}+\mu _{\rm m}\right)+\mu _{\rm e} \mu _{\rm m}\right)\right)\\
		4 \rho  \left(\kappa _{\rm e}+\kappa _{\rm m}\right) \left(\mu _{\rm e}+\mu _{\rm m}\right)\\
		4 \rho  L_{\rm c}^2 \left(\left(\bar{\kappa }_{\gamma}+\gammabar_1\right) \left(\kappa _{\rm e}+\kappa _{\rm m}\right) \left(\mu _{\rm e}+\mu _{\rm m}\right)+\gamma_1 \left(\mu _{\rm e} \kappa _{\rm m}+\kappa _{\rm e} \left(\mu _{\rm e}+\kappa _{\rm m}\right)\right)+\kappa _{\gamma} \left(\kappa _{\rm e} \left(\mu _{\rm e}+\mu _{\rm m}\right)+\mu _{\rm e} \mu _{\rm m}\right)\right)\\
		4 \rho ^2 L_{\rm c}^2 \left(\kappa _{\gamma} \left(\mu _{\rm e}+\mu _{\rm m}\right)+\gamma_1 \left(\kappa _{\rm e}+\kappa _{\rm m}\right)\right)\\
		4 \rho ^2 L_{\rm c}^4 \left(\kappa _{\gamma} \left(\bar{\kappa }_{\gamma}+\gammabar_1\right) \left(\mu _{\rm e}+\mu _{\rm m}\right)+\gamma_1 \left(\left(\bar{\kappa }_{\gamma}+\gammabar_1\right) \left(\kappa _{\rm e}+\kappa _{\rm m}\right)+\kappa _{\gamma} \left(\kappa _{\rm e}+\mu _{\rm e}\right)\right)\right)\\
		4 \gamma_1 \rho ^3 L_{\rm c}^4 \kappa _{\gamma} \\
		4 \gamma_1 \rho ^3 L_{\rm c}^6 \kappa _{\gamma} \left(\bar{\kappa }_{\gamma}+\gammabar_1\right)}
\end{equation}
Shear $0^\circ$:
\begin{equation}
	\matr{
		4 \mu _{\rm c} \mu _{\rm e}^* \mu _{\rm m}^*\\
		4 \rho  \mu _{\rm c} \left(\mu _{\rm e}^*+\mu _{\rm m}^*\right)\\
		4 \rho  L_{\rm c}^2 \left(\mu _{\rm c} \left(\gammabar_2 \left(\mu _{\rm e}^*+\mu _{\rm m}^*\right)+\gammabar_1^* \left(\mu _{\rm e}^*+\mu _{\rm m}^*\right)+\gamma_1^* \mu _{\rm e}^*\right)+\gamma_2 \left(\mu _{\rm c} \left(\mu _{\rm e}^*+\mu _{\rm m}^*\right)+\mu _{\rm e}^* \mu _{\rm m}^*\right)\right)\\
		4 \rho ^2 L_{\rm c}^2 \left(\mu _{\rm c} \gamma_1^*+\gamma_2 \left(\mu _{\rm e}^*+\mu _{\rm m}^*\right)\right)\\
		4 \rho ^2 L_{\rm c}^4 \left(\gammabar_2 \mu _{\rm c} \gamma_1^*+\mu _{\rm c} \gamma_1^* \gammabar_1^*+\gamma_2 \gammabar_2 \left(\mu _{\rm e}^*+\mu _{\rm m}^*\right)+\gamma_2 \gammabar_1^* \left(\mu _{\rm e}^*+\mu _{\rm m}^*\right)+\gamma_2 \gamma_1^* \left(\mu _{\rm e}^*+\mu _{\rm c}\right)\right)\\
		4 \gamma_2 \rho ^3 L_{\rm c}^4 \gamma_1^*\\
		4 \gamma_2 \rho ^3 L_{\rm c}^6 \gamma_1^* \left(\gammabar_1^*+\gammabar_2\right)}
\end{equation}
Pressure $45^\circ$:
\begin{equation}
	\matr{
		4 \left(\kappa _{\rm m} \mu _{\rm e}^* \mu _{\rm m}^*+\kappa _{\rm e} \left(\kappa _{\rm m} \left(\mu _{\rm e}^*+\mu _{\rm m}^*\right)+\mu _{\rm e}^* \mu _{\rm m}^*\right)\right)\\
		4 \rho  \left(\kappa _{\rm e}+\kappa _{\rm m}\right) \left(\mu _{\rm e}^*+\mu _{\rm m}^*\right) \\
        4 L_{\rm c}^2 \rho ((\kappa_{\rm e}+\kappa_{\rm m}) (\gamma_1^* \mu_{\rm e}^*+\gammabar_1^* (\mu_{\rm e}^*+\mu_{\rm m}^*)+\kappabar_\gamma \mu_{\rm e}^*+\kappabar_\gamma \mu_{\rm m}^*)+\gamma_1^* \kappa_{\rm e} \kappa_{\rm m}+\kappa \gamma  \mu_{\rm e}^* \mu_{\rm m}^*+\kappa \gamma  \kappa_{\rm e} (\mu_{\rm e}^*+\mu_{\rm m}^*))\\
		4 \rho ^2 L_{\rm c}^2 \left(\kappa _{\gamma} \left(\mu _{\rm e}^*+\mu _{\rm m}^*\right)+\gamma_1^* \left(\kappa _{\rm e}+\kappa _{\rm m}\right)\right) \\
		4 \rho ^2 L_{\rm c}^4 \left(\kappa _{\gamma} \left(\bar{\kappa }_{\gamma} \left(\mu _{\rm e}^*+\mu _{\rm m}^*\right)+\gammabar_1^* \mu _{\rm e}^*+\gammabar_1^* \mu _{\rm m}^*+\kappa _{\rm e} \gamma_1^*+\gamma_1^* \mu _{\rm e}^*\right)+\gamma_1^* \left(\kappa _{\rm e}+\kappa _{\rm m}\right) \left(\gammabar_1^*+\bar{\kappa }_{\gamma}\right)\right) \\
		4 \rho ^3 L_{\rm c}^4 \kappa _{\gamma} \gamma_1^* \\
		4 \rho ^3 L_{\rm c}^6 \kappa _{\gamma} \gamma_1^* \left(\gammabar_1^*+\bar{\kappa }_{\gamma}\right)}
\end{equation}
Shear $45^\circ$:
\begin{equation}
	\matr{
		4 \mu _{\rm c} \mu _{\rm e} \mu _{\rm m} \\
		4 \rho  \mu _{\rm c} \left(\mu _{\rm e}+\mu _{\rm m}\right) \\
		4 \rho  L_{\rm c}^2 \left(\left(\gammabar_1+\gammabar_2\right) \mu _{\rm c} \left(\mu _{\rm e}+\mu _{\rm m}\right)+\gamma_1 \mu _{\rm c} \mu _{\rm e}+\gamma_2 \left(\mu _{\rm c} \left(\mu _{\rm e}+\mu _{\rm m}\right)+\mu _{\rm e} \mu _{\rm m}\right)\right) \\
		4 \rho ^2 L_{\rm c}^2 \left(\gamma_1 \mu _{\rm c}+\gamma_2 \left(\mu _{\rm e}+\mu _{\rm m}\right)\right) \\
		4 \rho ^2 L_{\rm c}^4 \left(\gamma_1 \left(\left(\gammabar_1+\gammabar_2\right) \mu _{\rm c}+\gamma_2 \left(\mu _{\rm c}+\mu _{\rm e}\right)\right)+\gamma_2 \left(\gammabar_1+\gammabar_2\right) \left(\mu _{\rm e}+\mu _{\rm m}\right)\right) \\
		4 \gamma_1 \gamma_2 \rho ^3 L_{\rm c}^4 \\
		4 \gamma_1 \gamma_2 \rho ^3 \left(\gammabar_1+\gammabar_2\right) L_{\rm c}^6}
\end{equation}
%
%
%
%
%

\section{Coefficients for the dispersion curves with $\Curl P$}\label{app:coefficients2}

For the relaxed micromorphic model with $\Curl P$ but without $\Curl\dot P$, we arrived \eqref{eq:dispersionRelationCurlP} at the dispersion polynomial
\begin{equation}
	c_0\.k^2+c_1\.k^4-(c_2+c_3\.k^2+c_4^*\.k^4)\,\omega^2+(c_5+c_6^*\.k^2+c_7^*\.k^4)\,\omega^4-(c_8+c_9^star\,k^2)\,\omega^6
\end{equation}
The coefficients $c_0,c_2,c_5,c_8,c_9$ are identical to the one given in the section above without $\Curl P$. Thus we list $(c_1,c_3,c_4^*,c_6^*,c_7^*)^T$ using $c_3^*$ and $c_6^*$ to shorten some expressions.

Pressure $0^\circ$:
\begin{equation}
	\matr{
		2\alpha_1 L_{\rm c}^2 \left(\kappa _{\rm e}+\mu _{\rm e}\right) \left(\kappa _{\rm m}+\mu _{\rm m}\right) \\
		c_3^*+2\rho  \alpha_1 L_{\rm c}^2 \left(\kappa _{\rm e}+\mu _{\rm e}+\kappa _{\rm m}+\mu _{\rm m}\right) \\
		2\rho  \alpha_1 L_{\rm c}^4 \left(\gammabar_1 \left(\kappa _{\rm e}+\mu _{\rm e}+\kappa _{\rm m}+\mu _{\rm m}\right)+\bar{\kappa }_{\gamma} \left(\kappa _{\rm e}+\mu _{\rm e}+\kappa _{\rm m}+\mu _{\rm m}\right)+\left(\kappa _{\gamma}+\gamma_1\right) \left(\kappa _{\rm e}+\mu _{\rm e}\right)\right) \\
		c_6^*+2\rho ^2 \alpha_1 L_{\rm c}^4 \left(\kappa _{\gamma}+\gamma_1\right) \\
		2\rho ^2 \alpha_1 L_{\rm c}^6 \left(\kappa _{\gamma}+\gamma_1\right) \left(\bar{\kappa }_{\gamma}+\gammabar_1\right)}
\end{equation}
Shear $0^\circ$:
\begin{equation}
	\matr{
		2\alpha_1 L_{\rm c}^2 \mu _{\rm m}^* \left(\mu _{\rm e}^*+\mu _{\rm c}\right) \\
		c_3^*+2\rho  \alpha_1 L_{\rm c}^2 \left(\mu _{\rm e}^*+\mu _{\rm m}^*+\mu _{\rm c}\right) \\
		2\rho  \alpha_1 L_{\rm c}^4 \left(\gammabar_2 \left(\mu _{\rm e}^*+\mu _{\rm m}^*+\mu _{\rm c}\right)+\gammabar_1^* \left(\mu _{\rm e}^*+\mu _{\rm m}^*+\mu _{\rm c}\right)+\left(\gamma_1^*+\gamma_2\right) \left(\mu _{\rm e}^*+\mu _{\rm c}\right)\right) \\
		c_6^*+2\rho ^2 \alpha_1 L_{\rm c}^4 \left(\gamma_1^*+\gamma_2\right) \\
		2\rho ^2 \alpha_1 L_{\rm c}^6 \left(\gamma_1^*+\gamma_2\right) \left(\gammabar_1^*+\gammabar_2\right)}
\end{equation}
Pressure $45^\circ$:
\begin{equation}
	\matr{
		2\alpha_1 L_{\rm c}^2 \left(\mu _{\rm e}^*+\kappa _{\rm e}\right) \left(\mu _{\rm m}^*+\kappa _{\rm m}\right) \\
		c_3^*+2\rho  \alpha_1 L_{\rm c}^2 \left(\mu _{\rm e}^*+\mu _{\rm m}^*+\kappa _{\rm e}+\kappa _{\rm m}\right) \\
		2\rho  \alpha_1 L_{\rm c}^4 \left(\bar{\kappa }_{\gamma} \left(\mu _{\rm e}^*+\mu _{\rm m}^*+\kappa _{\rm e}+\kappa _{\rm m}\right)+\gammabar_1^* \left(\mu _{\rm e}^*+\mu _{\rm m}^*+\kappa _{\rm e}+\kappa _{\rm m}\right)+\left(\gamma_1^*+\kappa _{\gamma}\right) \left(\mu _{\rm e}^*+\kappa _{\rm e}\right)\right) \\
		c_6^*+2\rho ^2 \alpha_1 L_{\rm c}^4 \left(\gamma_1^*+\kappa _{\gamma}\right) \\
		2\rho ^2 \alpha_1 L_{\rm c}^6 \left(\gamma_1^*+\kappa _{\gamma}\right) \left(\gammabar_1^*+\bar{\kappa }_{\gamma}\right)}
\end{equation}
Shear $45^\circ$:
\begin{equation}
	\matr{
		2\alpha_1 L_{\rm c}^2 \mu _{\rm m} \left(\mu _{\rm c}+\mu _{\rm e}\right) \\
		c_3^*+2\rho  \alpha_1 L_{\rm c}^2 \left(\mu _{\rm c}+\mu _{\rm e}+\mu _{\rm m}\right) \\
		2\rho  \alpha_1 L_{\rm c}^4 \left(\gammabar_1 \left(\mu _{\rm c}+\mu _{\rm e}+\mu _{\rm m}\right)+\gammabar_2 \left(\mu _{\rm c}+\mu _{\rm e}+\mu _{\rm m}\right)+\left(\gamma_1+\gamma_2\right) \left(\mu _{\rm c}+\mu _{\rm e}\right)\right) \\
		c_6^*+2\left(\gamma_1+\gamma_2\right) \rho ^2 \alpha_1 L_{\rm c}^4 \\
		2\left(\gamma_1+\gamma_2\right) \rho ^2 \left(\gammabar_1+\gammabar_2\right) \alpha_1 L_{\rm c}^6}
\end{equation}
%
%
%
%
%

\section{Coefficients for the dispersion curves with $\Curl\dot P$}\label{app:coefficients3}
For the relaxed micromorphic model with curvature, we arrived \eqref{eq:dispersionRelationCurldotP} at the dispersion polynomial
\begin{equation}
	c_0\.k^2+c_1\.k^4-(c_2+c_3\.k^2+c_4\.k^4)\,\omega^2+(c_5+c_6\.k^2+c_7\.k^4)\,\omega^4-(c_8+c_9\,k^2+c_{10}\,k^4)\,\omega^6
\end{equation}
The coefficients $c_0,c_1,c_2,c_3,c_5,c_6,c_8$ are identical to the one given in the section above without $\Curl\dot P$. Thus we list $(c_4,c_6,c_7,c_9,c_{10})^T$ using $c_4^*,c_6^*,c_7^*,c_9^*$ as the coefficients from previous sections above to shorten the expressions.

Pressure $0^\circ$:
\begin{equation}
	\matr{
		c_4^*+ 2\rho  \beta_1 L_{\rm c}^4 \left(\kappa _{\rm e}+\mu _{\rm e}\right) \left(\kappa _{\rm m}+\mu _{\rm m}\right) \\
		c_6^*+2 \rho ^2 \beta_1 L_{\rm c}^4 \left(\kappa _{\rm e}+\mu _{\rm e}+\kappa _{\rm m}+\mu _{\rm m}\right) \\
		c_7^*+2 \rho ^2 \beta_1 L_{\rm c}^6 \left(\gammabar_1 \left(\kappa _{\rm e}+\mu _{\rm e}+\kappa _{\rm m}+\mu _{\rm m}\right)+\bar{\kappa }_{\gamma} \left(\kappa _{\rm e}+\mu _{\rm e}+\kappa _{\rm m}+\mu _{\rm m}\right)+\left(\kappa _{\gamma}+\gamma_1\right) \left(\kappa _{\rm e}+\mu _{\rm e}\right)\right) \\
		c_9^*+2 \rho ^3 \beta_1 L_{\rm c}^6 \left(\kappa _{\gamma}+\gamma_1\right) \\
		2\rho ^3 \beta_1 L_{\rm c}^8 \left(\kappa _{\gamma}+\gamma_1\right) \left(\bar{\kappa }_{\gamma}+\gammabar_1\right)}
\end{equation}
Shear $0^\circ$:
\begin{equation}
	\matr{
		c_4^*+2 \rho  \beta_1 L_{\rm c}^4 \mu _{\rm m}^* \left(\mu _{\rm e}^*+\mu _{\rm c}\right) \\
		c_6^*+2 \rho ^2 \beta_1 L_{\rm c}^4 \left(\mu _{\rm e}^*+\mu _{\rm m}^*+\mu _{\rm c}\right) \\
		c_7^*+2 \rho ^2 \beta_1 L_{\rm c}^6 \left(\gammabar_2 \left(\mu _{\rm e}^*+\mu _{\rm m}^*+\mu _{\rm c}\right)+\gammabar_1^* \left(\mu _{\rm e}^*+\mu _{\rm m}^*+\mu _{\rm c}\right)+\left(\gamma_1^*+\gamma_2\right) \left(\mu _{\rm e}^*+\mu _{\rm c}\right)\right) \\
		c_9^*+ 2\rho ^3 \beta_1 L_{\rm c}^6 \left(\gamma_1^*+\gamma_2\right) \\
		2\rho ^3 \beta_1 L_{\rm c}^8 \left(\gamma_1^*+\gamma_2\right) \left(\gammabar_1^*+\gammabar_2\right)}
\end{equation}
Pressure $45^\circ$:
\begin{equation}
	\matr{
		c_4^*+2 \rho \beta_1 L_{\rm c}^4 \left(\mu _{\rm e}^*+\kappa _{\rm e}\right) \left(\mu _{\rm m}^*+\kappa _{\rm m}\right) \\
		c_6^*+2 \rho ^2 \beta_1 L_{\rm c}^4 \left(\mu _{\rm e}^*+\mu _{\rm m}^*+\kappa _{\rm e}+\kappa _{\rm m}\right) \\
		c_7^*+2 \rho ^2 \beta_1 L_{\rm c}^6 \left(\bar{\kappa }_{\gamma} \left(\mu _{\rm e}^*+\mu _{\rm m}^*+\kappa _{\rm e}+\kappa _{\rm m}\right)+\gammabar_1^* \left(\mu _{\rm e}^*+\mu _{\rm m}^*+\kappa _{\rm e}+\kappa _{\rm m}\right)+\left(\gamma_1^*+\kappa _{\gamma}\right) \left(\mu _{\rm e}^*+\kappa _{\rm e}\right)\right) \\
		c_9^*+ 2\rho ^3 \beta_1 L_{\rm c}^6 \left(\gamma_1^*+\kappa _{\gamma}\right) \\
		2\rho ^3 \beta_1 L_{\rm c}^8 \left(\gamma_1^*+\kappa _{\gamma}\right) \left(\gammabar_1^*+\bar{\kappa }_{\gamma}\right)}
\end{equation}
Shear $45^\circ$:
\begin{equation}
	\matr{
		c_4^*+ 2\rho  \beta_1 L_{\rm c}^4 \mu _{\rm m} \left(\mu _{\rm c}+\mu _{\rm e}\right) \\
		c_6^*+ 2\rho ^2 \beta_1 L_{\rm c}^4 \left(\mu _{\rm c}+\mu _{\rm e}+\mu _{\rm m}\right) \\
		c_7^*+2 \rho ^2 \beta_1 L_{\rm c}^6 \left(\gammabar_1 \left(\mu _{\rm c}+\mu _{\rm e}+\mu _{\rm m}\right)+\gammabar_2 \left(\mu _{\rm c}+\mu _{\rm e}+\mu _{\rm m}\right)+\left(\gamma_1+\gamma_2\right) \left(\mu _{\rm c}+\mu _{\rm e}\right)\right) \\
		c_9^*+ 2\left(\gamma_1+\gamma_2\right) \rho ^3 \beta_1 L_{\rm c}^6 \\
		2\left(\gamma_1+\gamma_2\right) \rho ^3 \left(\gammabar_1+\gammabar_2\right) \beta_1 L_{\rm c}^8}
\end{equation}
%
%
%
%
%
\end{document}